\documentclass[
 onecolumn, notitlepage,
superscriptaddress,
amsmath,amssymb,aps,
]{revtex4-1}

\usepackage{graphicx}
\usepackage{xcolor}
\usepackage{gensymb}
\usepackage{eso-pic}

\newcommand*\bu{\mathbf{u}}
\newcommand*\cm{c_{\text{m}}}
\newcommand*\cmIC{c_{\text{m},0}}
\newcommand*\cp{c_{\text{p}}}
\newcommand*\Dm{D_{\text{m}}}
\newcommand*\Dp{D_{\text{p}}}
\newcommand*\Ron{R_{\text{on}}}
\newcommand*\kappac{\kappa_{\text{c}}}

\newcommand*\koff{k_{\text{off}}}
\newcommand*\kd{k_{\text{d}}}

\newcommand{\eqn}{Eqn. }
\newcommand{\eqns}{Eqns. }

\newcommand\includefrontpdf[1]{%
  \clearpage
  \AddToShipoutPictureBG*{%
    \includegraphics[width=\paperwidth,height=\paperheight]{#1}%
  }%
  \thispagestyle{empty} 
  \null\vfill\clearpage
  \ClearShipoutPictureBG 
}

\begin{document}

\includefrontpdf{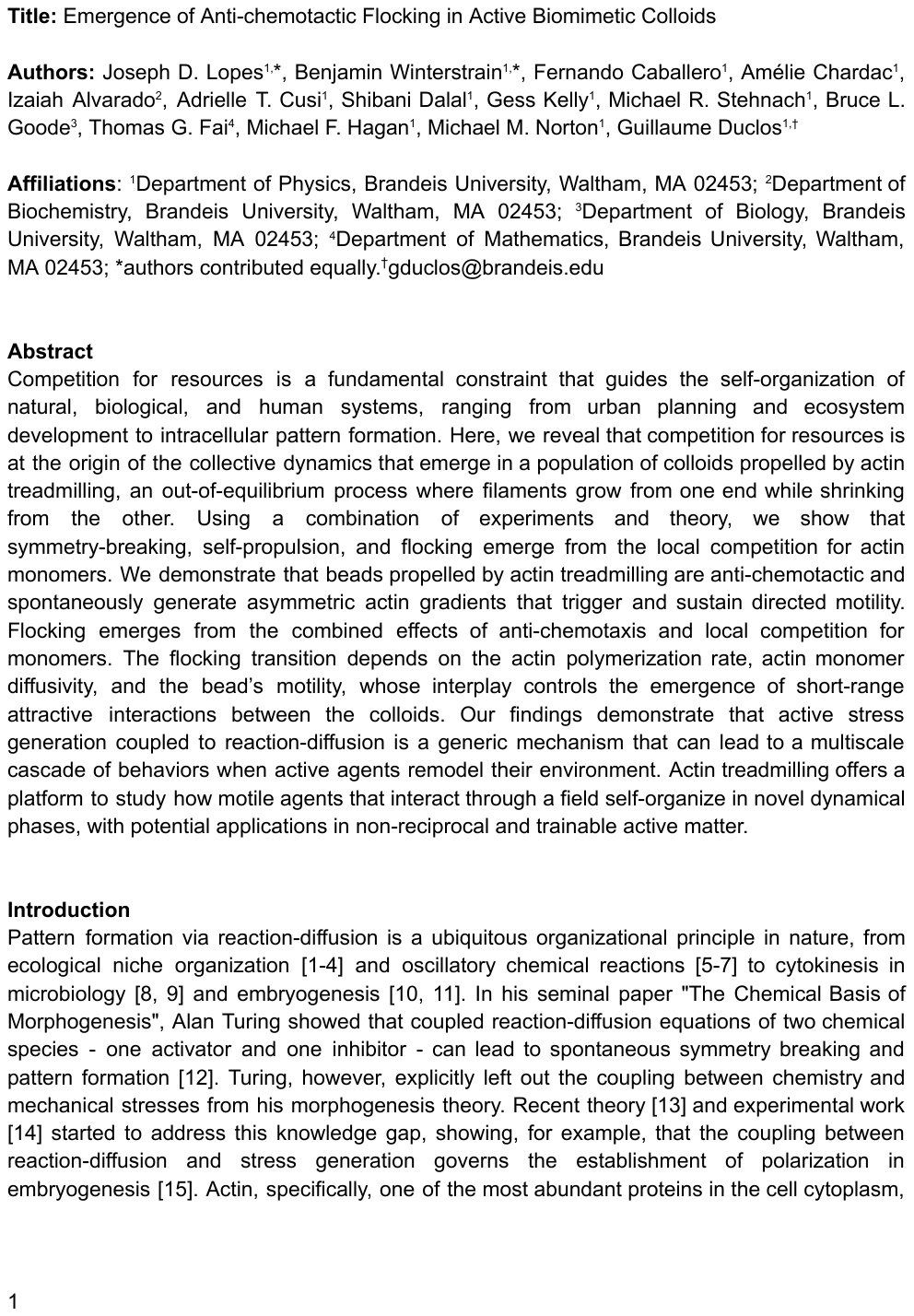}
\includefrontpdf{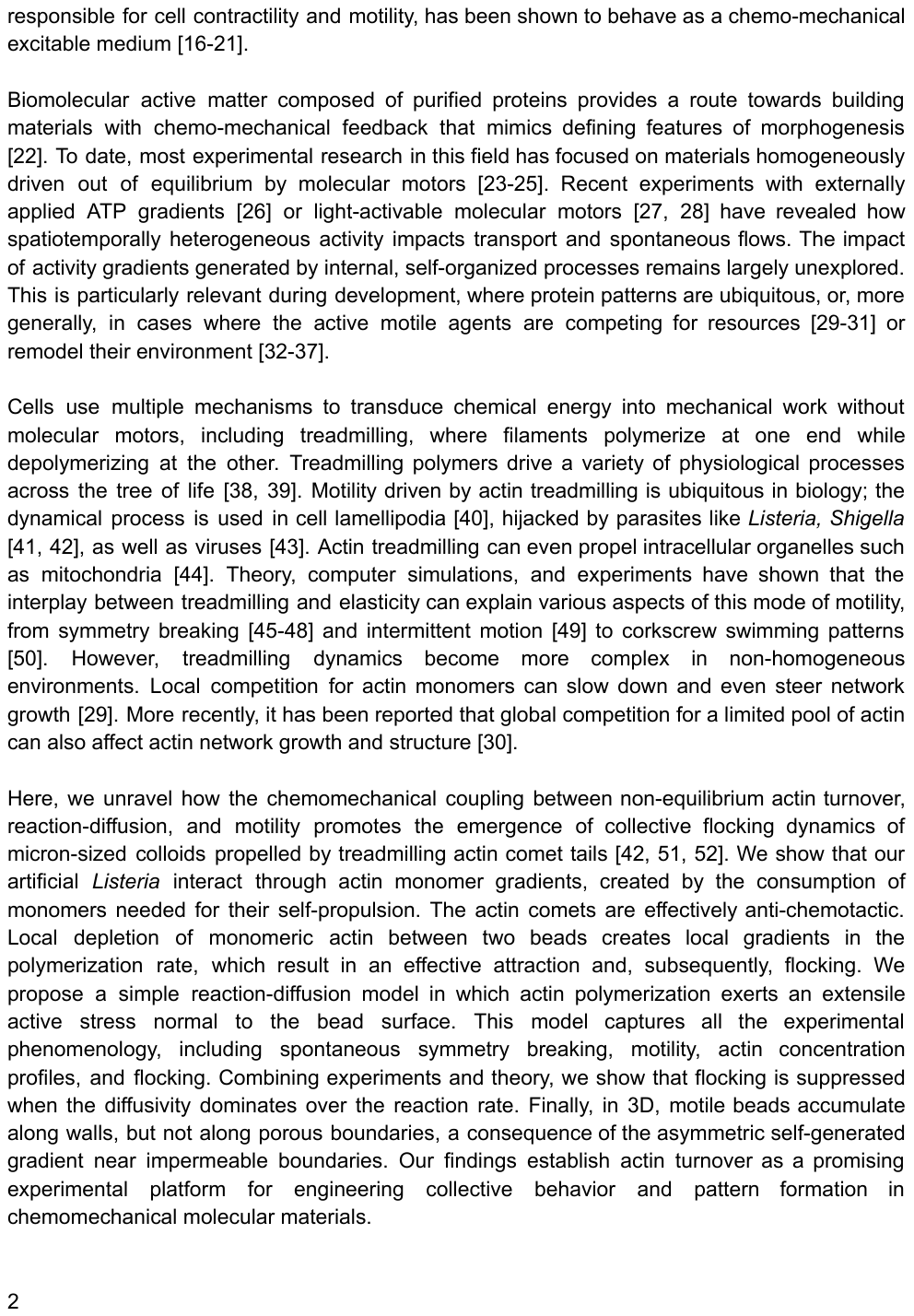}
\includefrontpdf{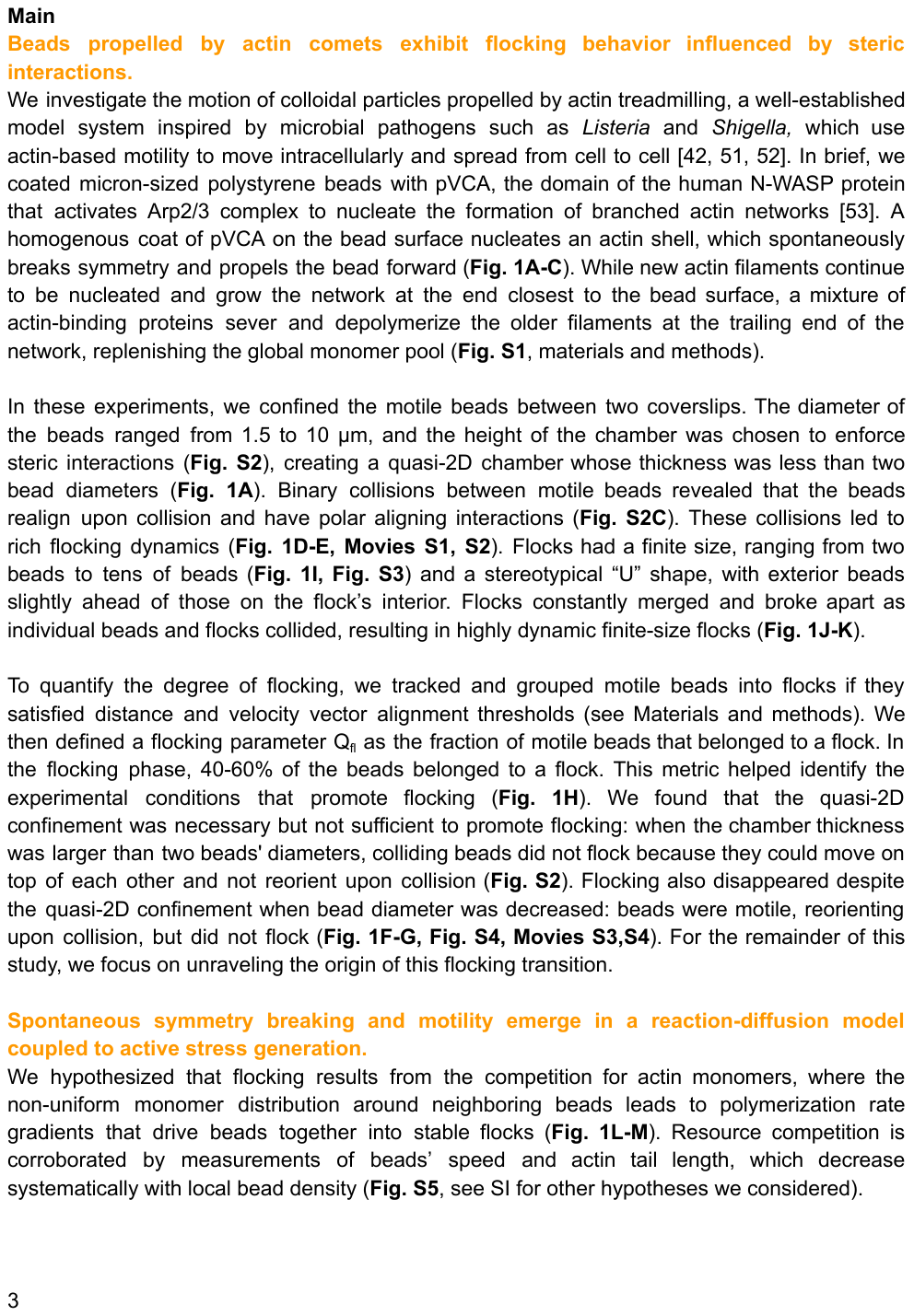}
\includefrontpdf{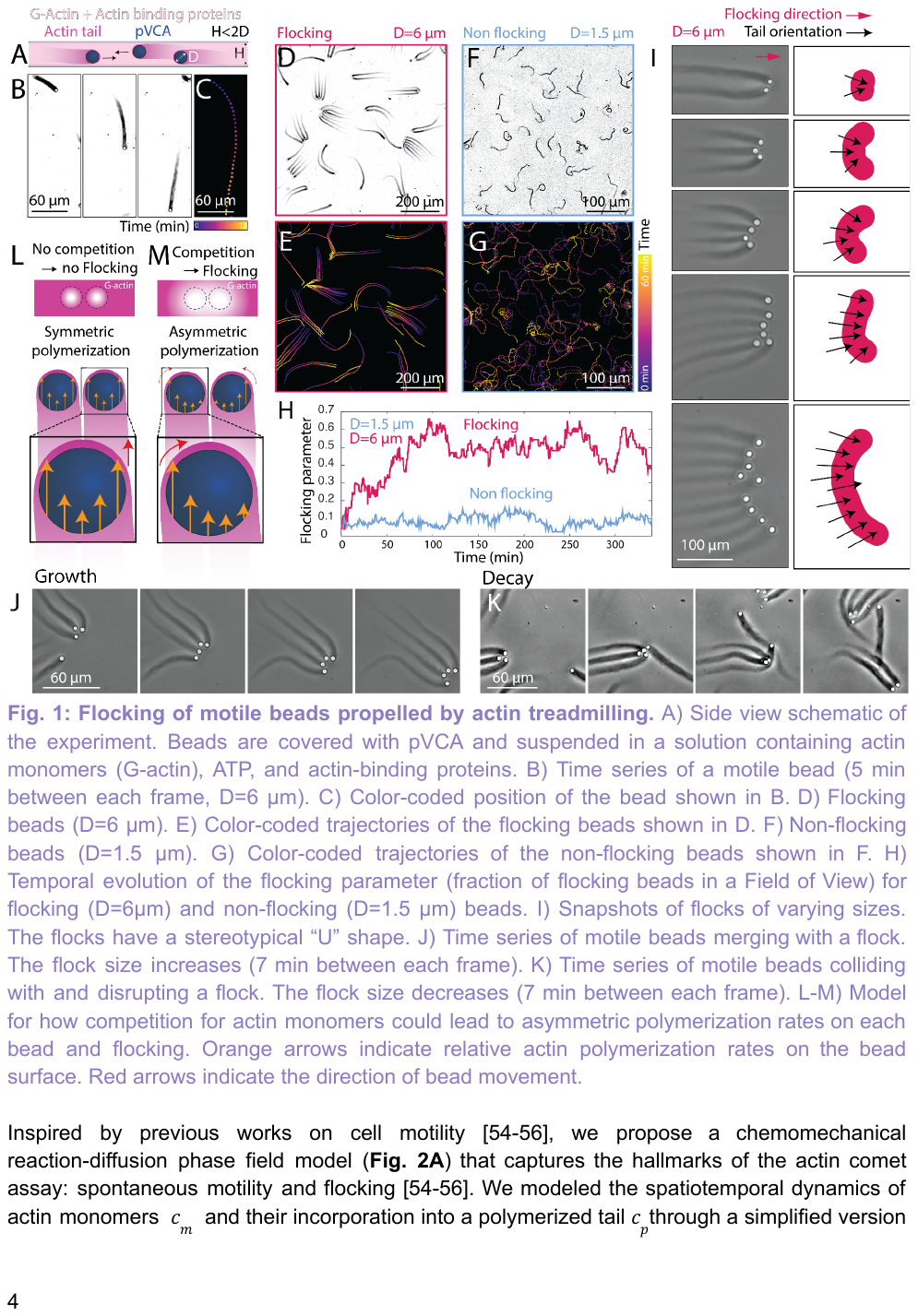}
\includefrontpdf{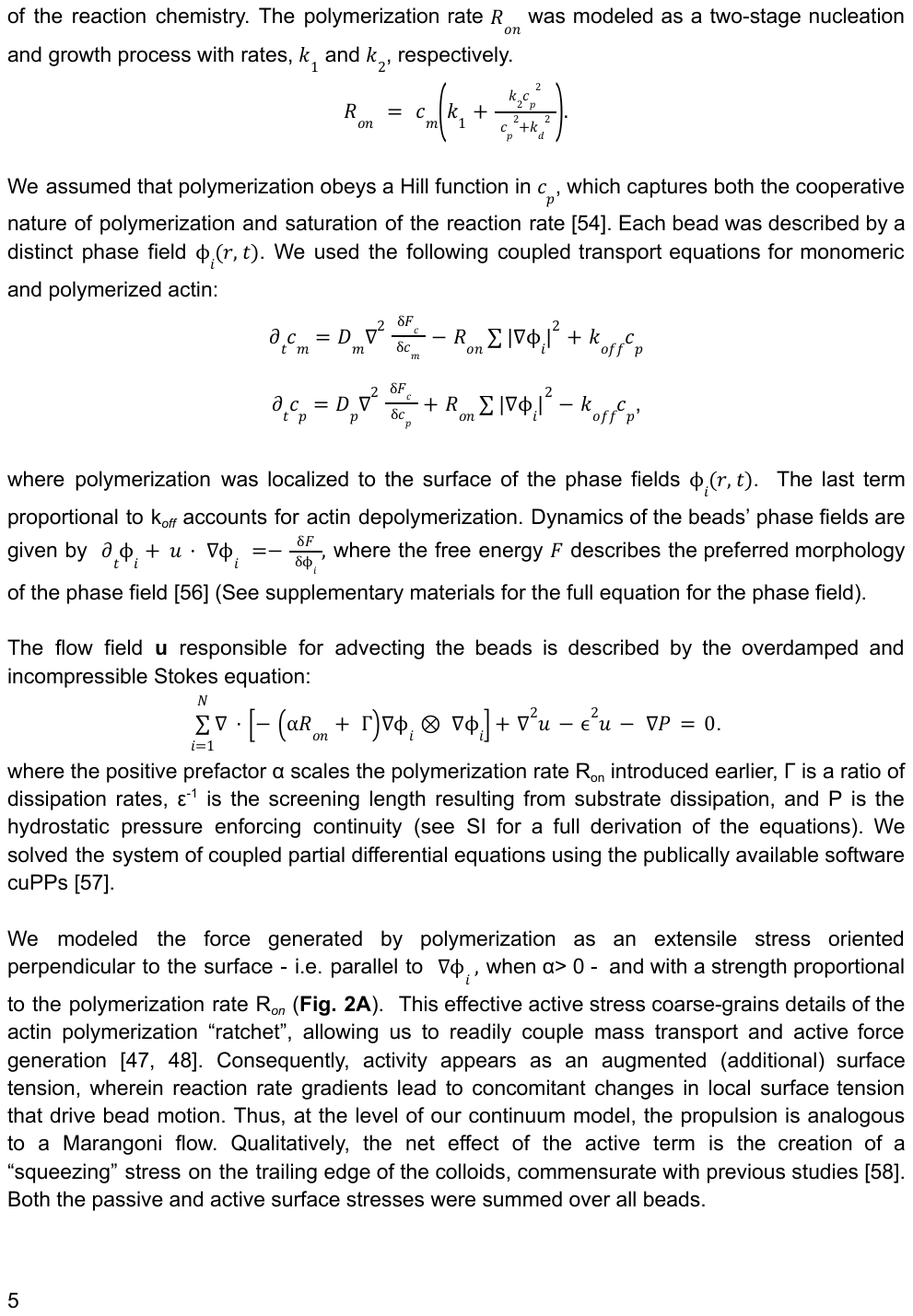}
\includefrontpdf{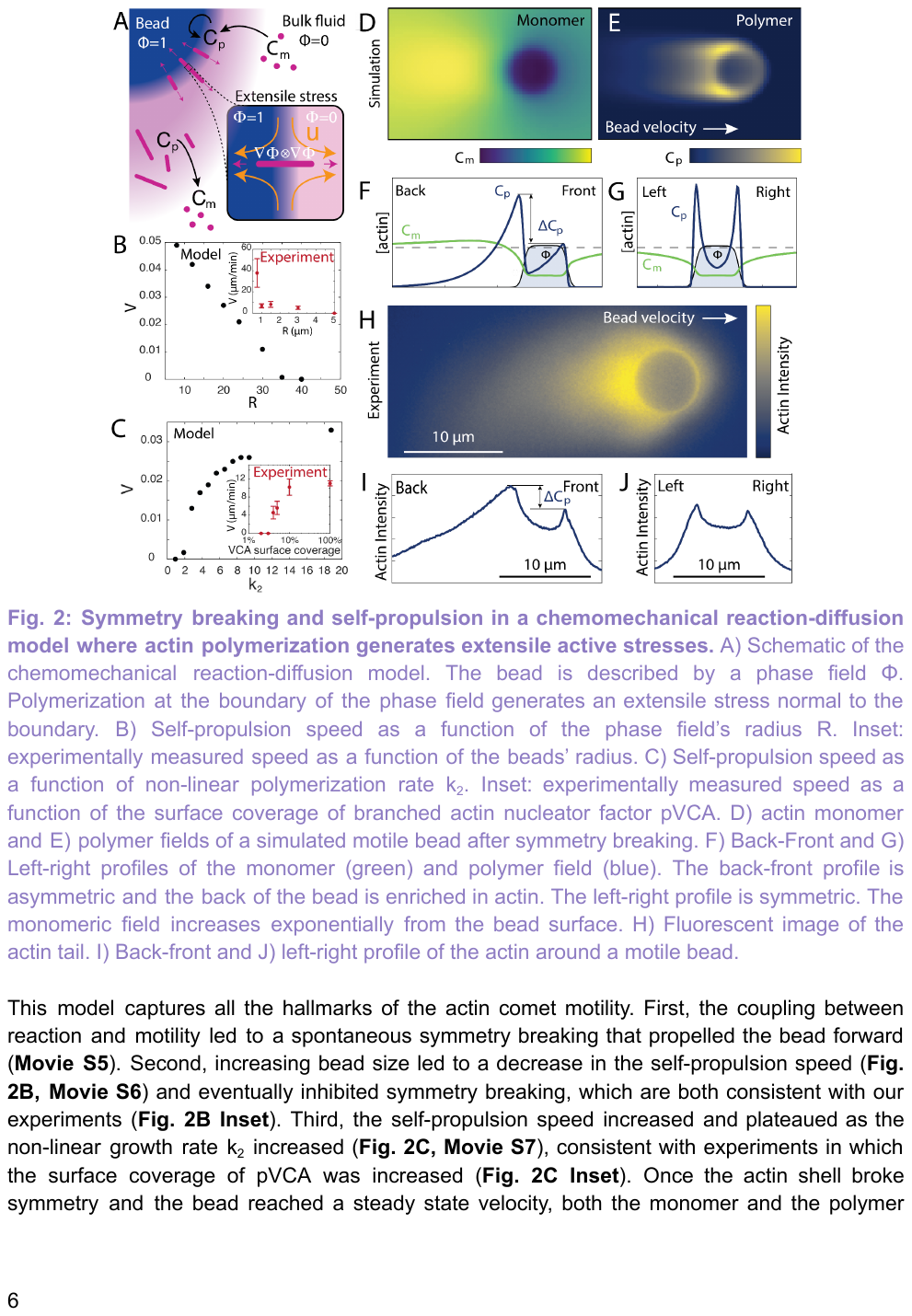}
\includefrontpdf{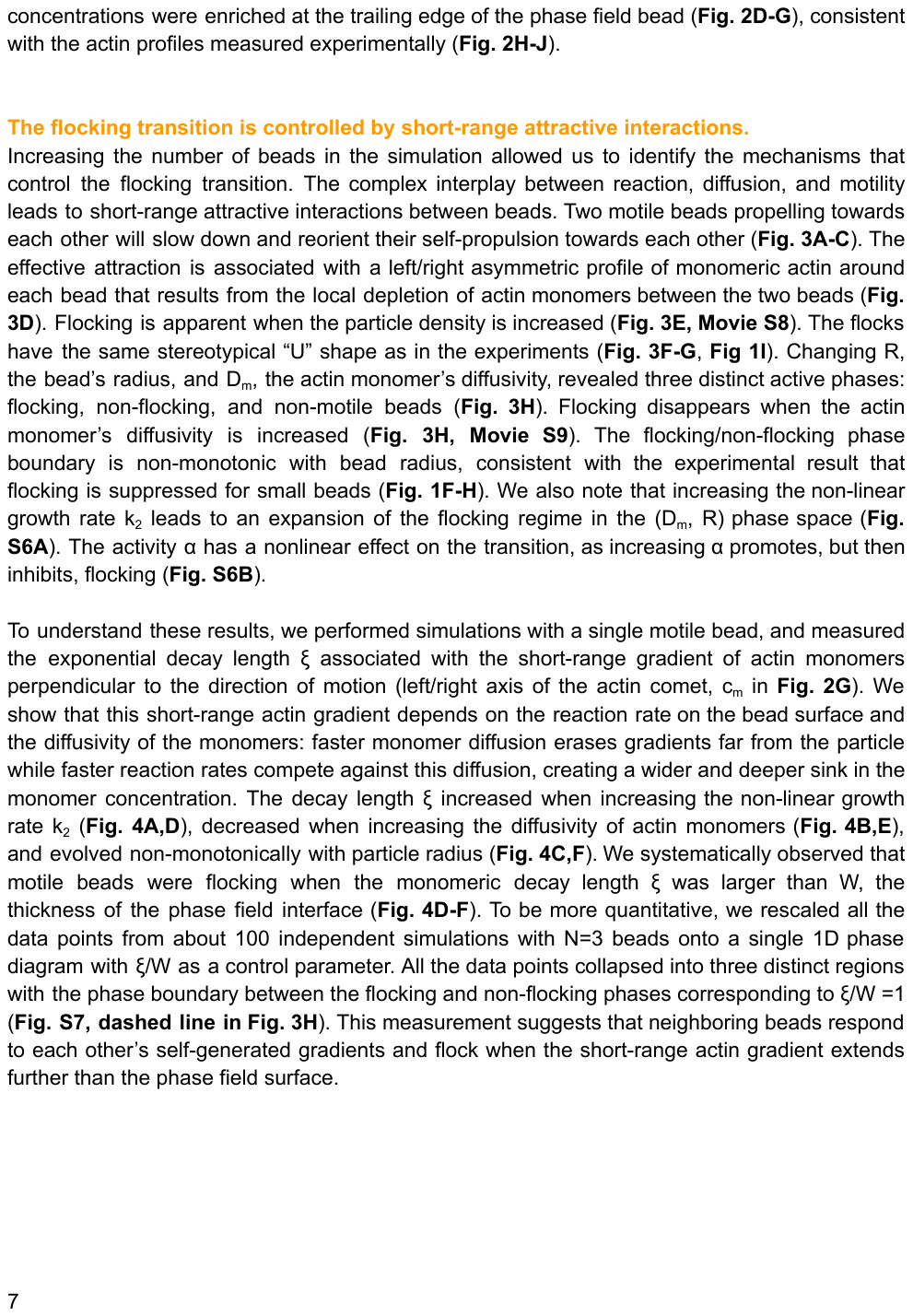}
\includefrontpdf{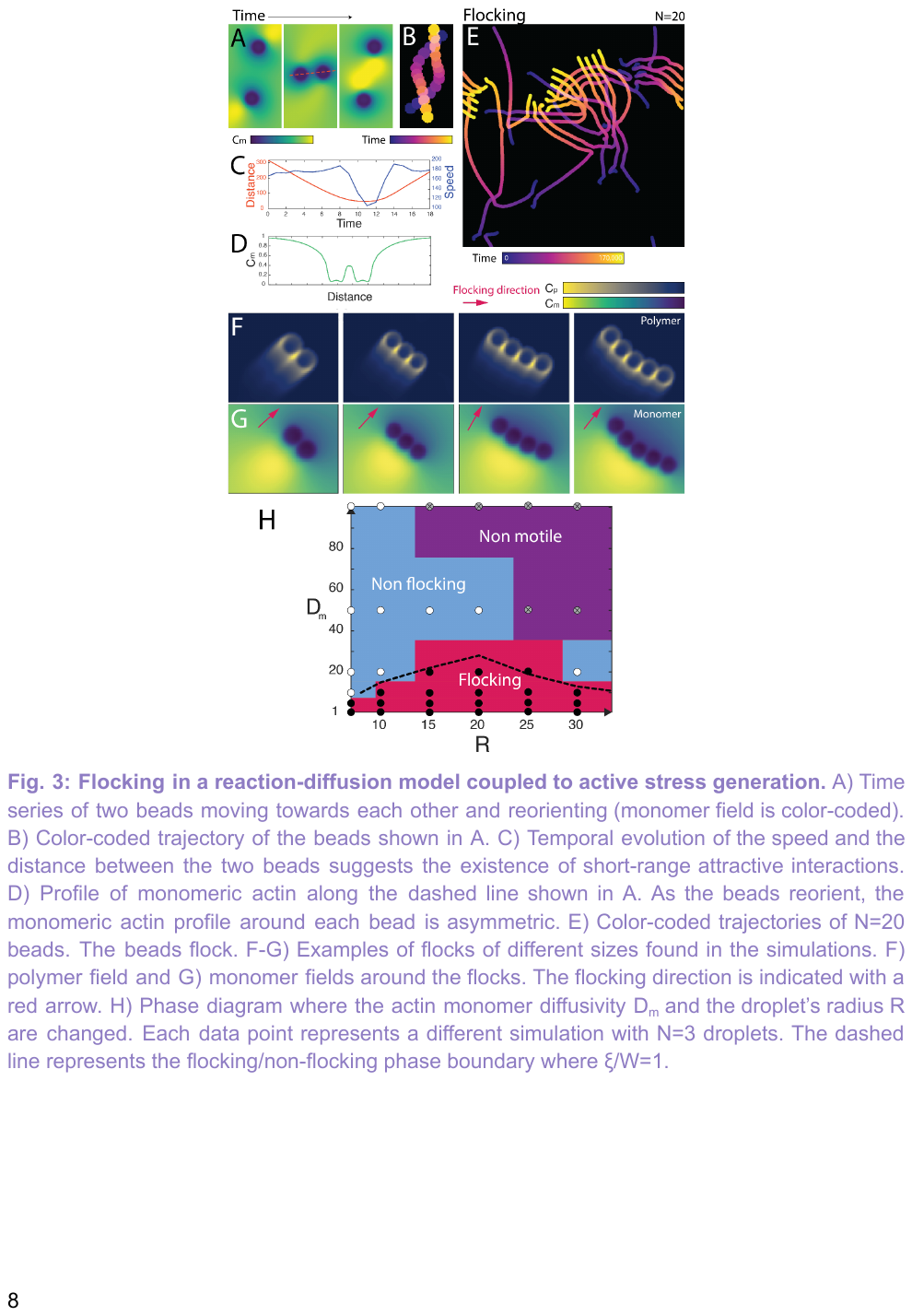}
\includefrontpdf{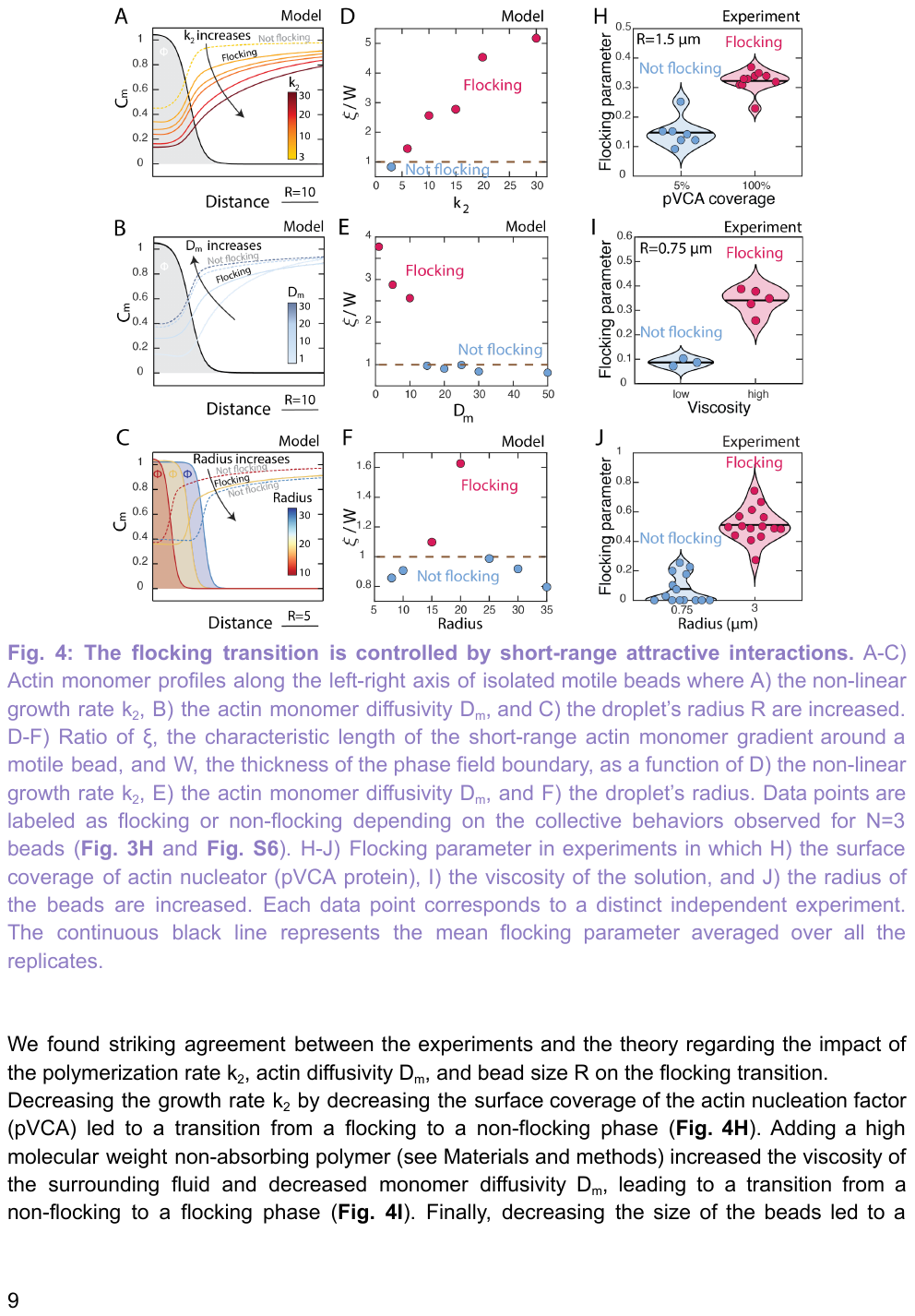}
\includefrontpdf{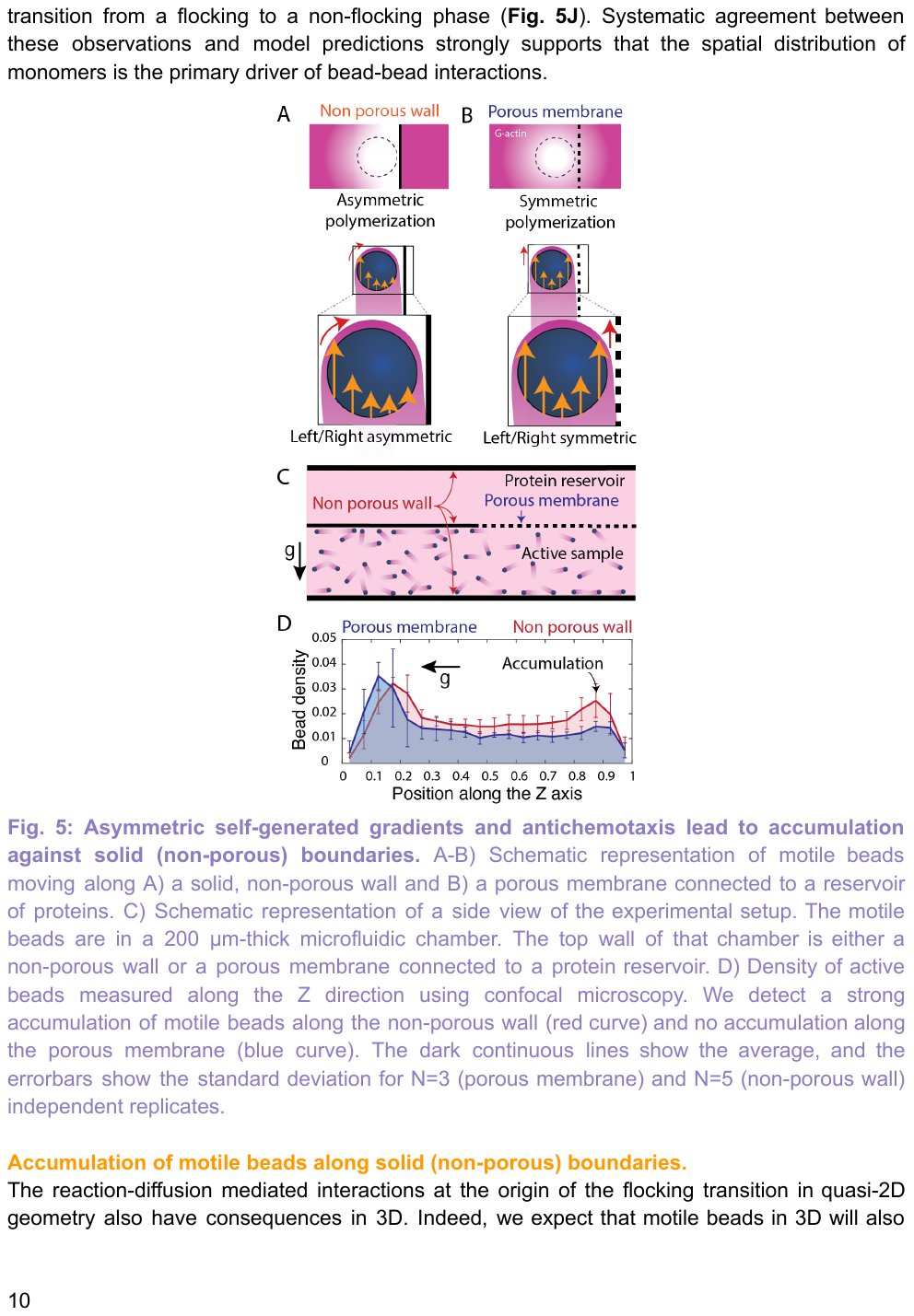}
\includefrontpdf{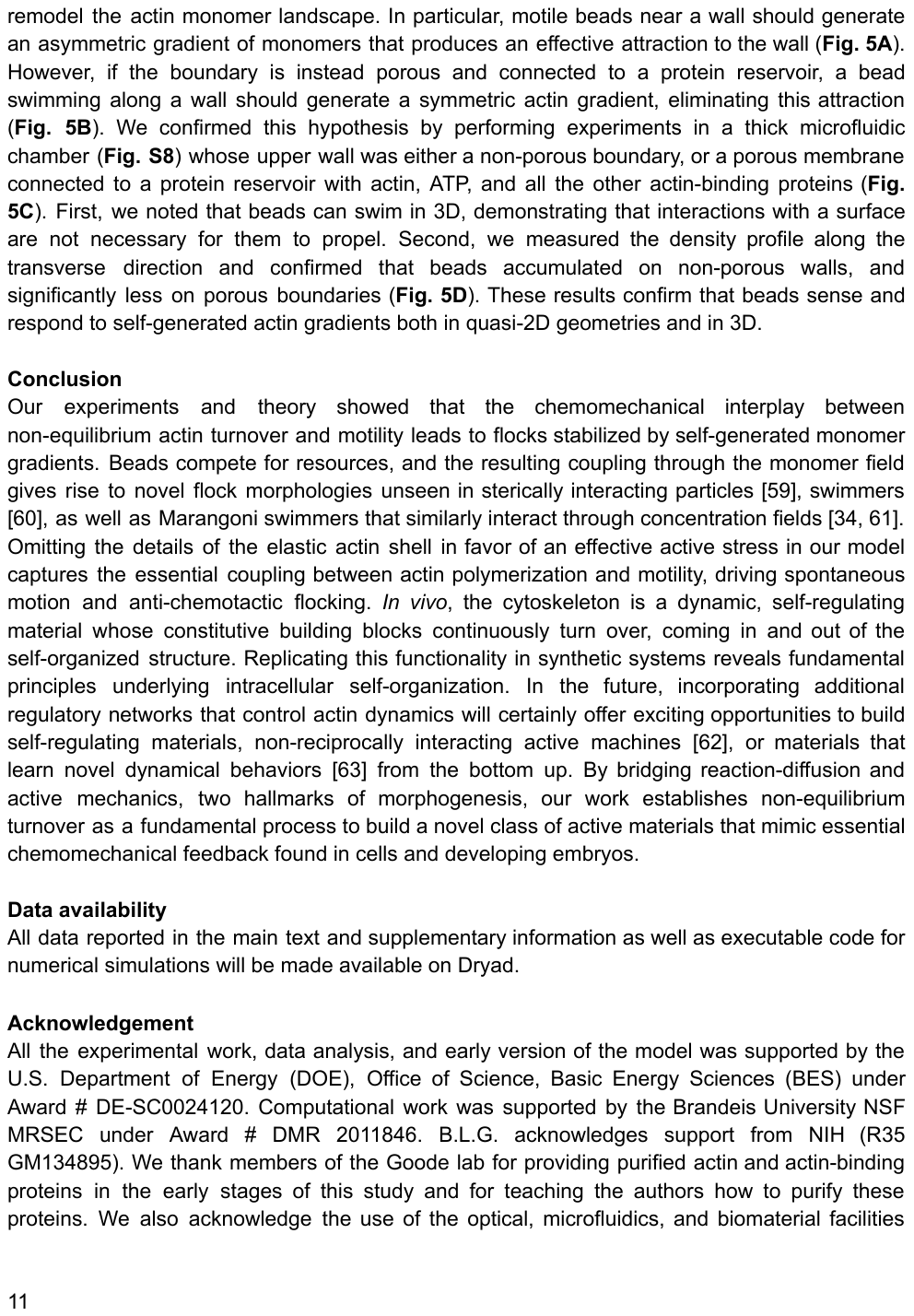}
\includefrontpdf{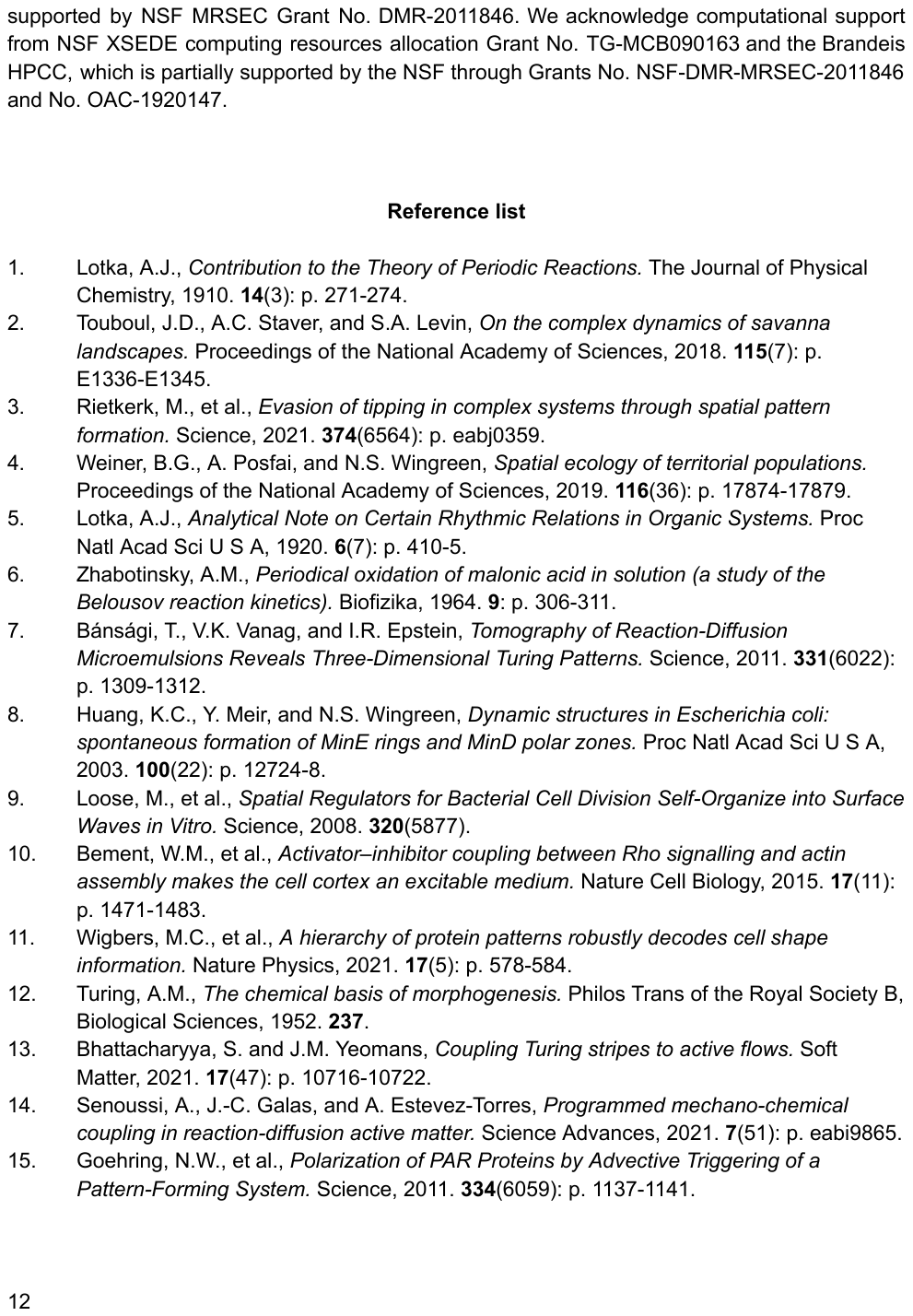}
\includefrontpdf{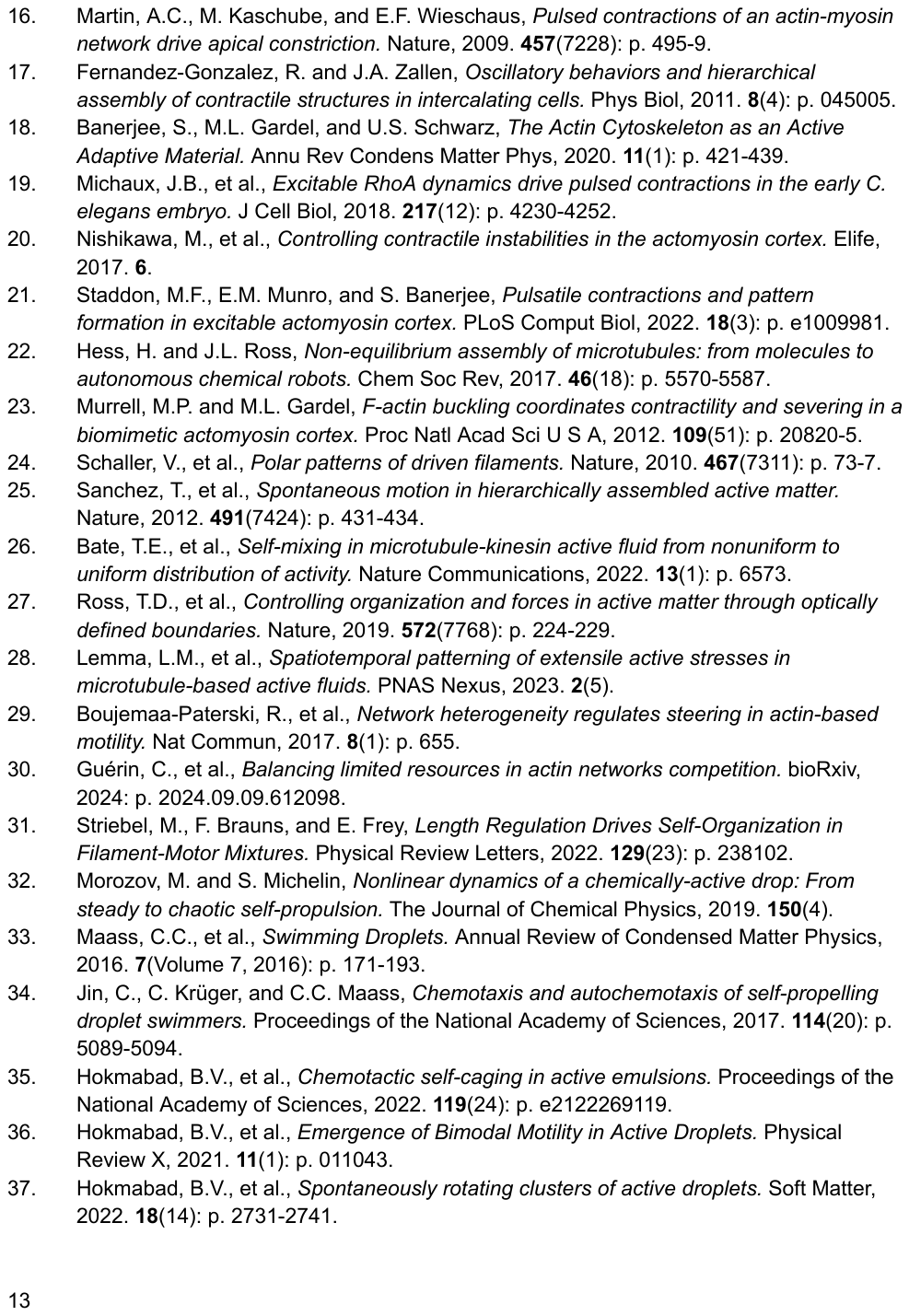}
\includefrontpdf{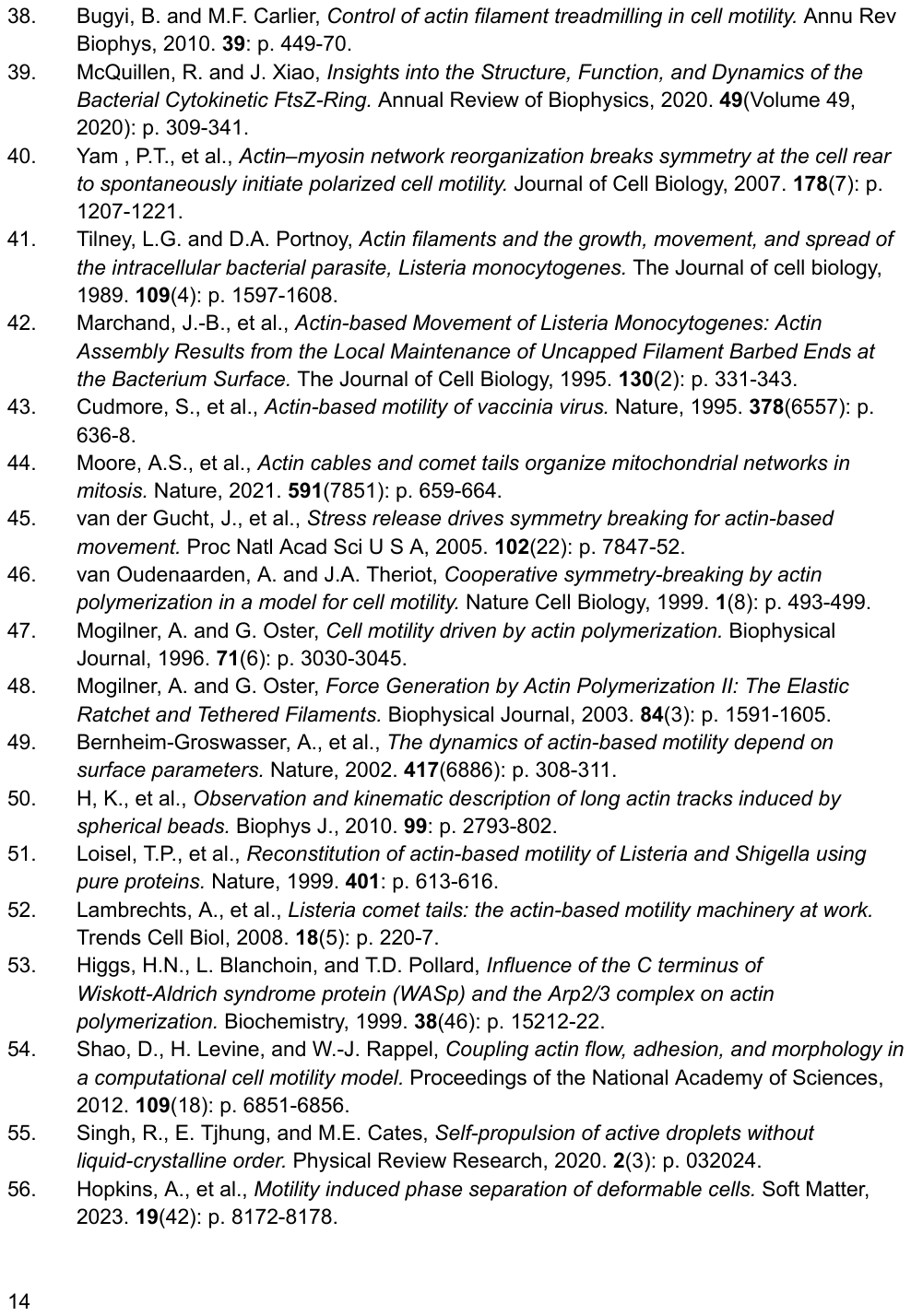}
\includefrontpdf{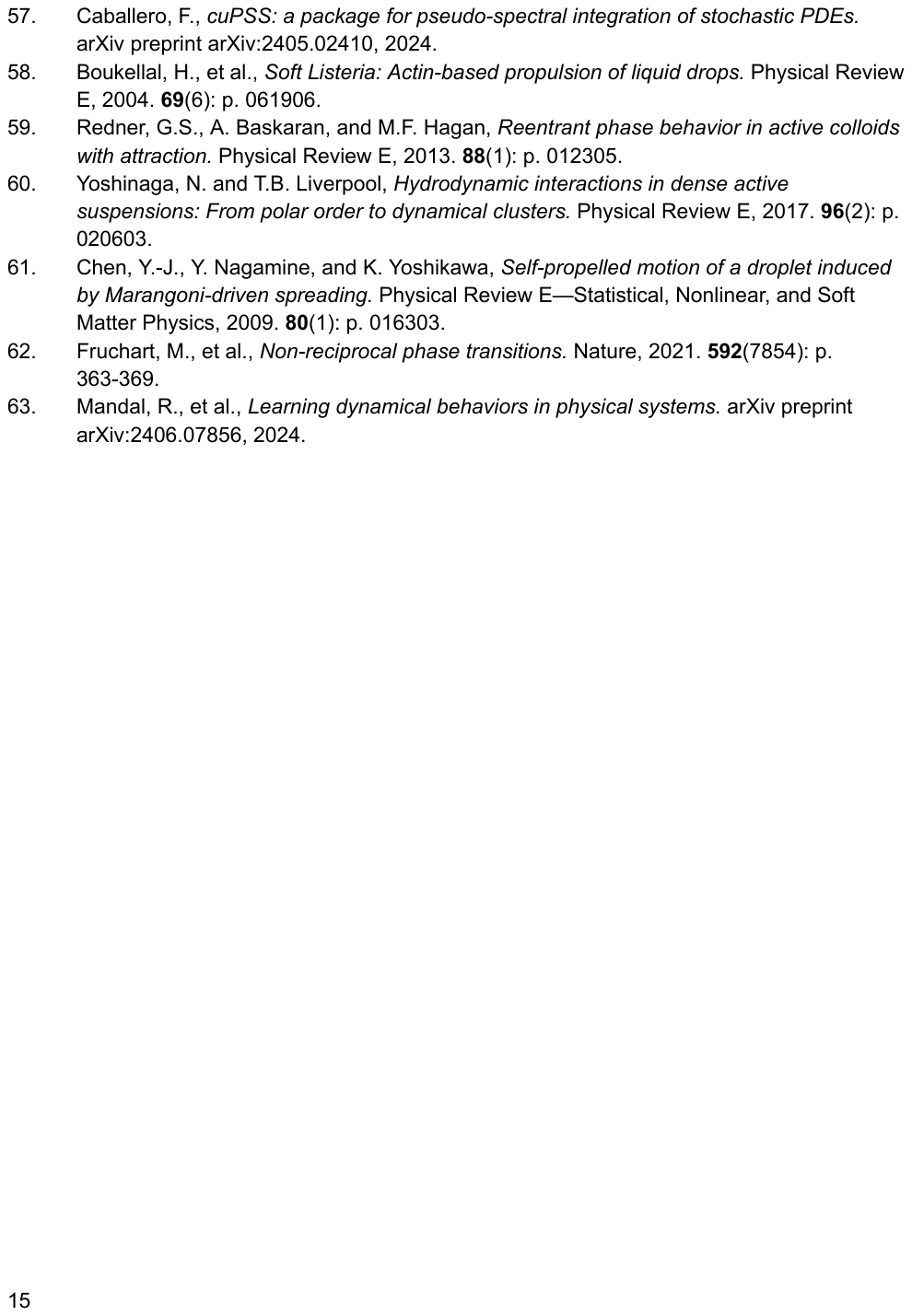}

\renewcommand{\thefigure}{S\arabic{figure}}
\renewcommand{\thetable}{S\arabic{table}}
\def\theequation{S\arabic{equation}}

\title{SUPPLEMENTARY INFORMATION \\~\\  Emergence of Anti-chemotactic Flocking in Active Biomimetic Colloids}
\author{Joseph D. Lopes*}
\author{Benjamin Winterstrain*}
\author{Fernando Caballero}
\author{Am\'elie Chardac}
\affiliation{Department of Physics, Brandeis University, 415 South street, Waltham, MA 02453}
\author{Izaiah Alvarado}
\affiliation{Department of Biochemistry, Brandeis University, 415 South street, Waltham, MA 02453}
\author{Adrielle T. Cusi}
\affiliation{Department of Physics, Brandeis University, 415 South street, Waltham, MA 02453}
\author{Shibani Dalal}
\affiliation{Department of Physics, Brandeis University, 415 South street, Waltham, MA 02453}
\author{Gess Kelly}
\affiliation{Department of Physics, Brandeis University, 415 South street, Waltham, MA 02453}
\author{Michael R. Stehnach}
\affiliation{Department of Physics, Brandeis University, 415 South street, Waltham, MA 02453}
\author{Bruce L. Goode}
\affiliation{Department of Biology, Brandeis University, 415 South street, Waltham, MA 02453}
\author{Thomas G. Fai}
\affiliation{Department of Mathematics, Brandeis University, 415 South street, Waltham, MA 02453}
\author{Michael F. Hagan}
\author{Michael M. Norton}
\author{Guillaume Duclos}
\affiliation{Department of Physics, Brandeis University, 415 South street, Waltham, MA 02453}


\maketitle

\section{Description of the Supplementary Movies}

\begin{itemize}
    \item{\textbf{Movie S1:} \textbf{Flocking phase.} Time series of 6 µm-diameter beads propelled by an actin comet (phase contrast microscopy, 15X magnification, total duration: 5h).}
    \item{\textbf{Movie S2:} \textbf{Binary collision in the flocking phase}: 6 µm-diameter beads are propelled by an actin comet (phase contrast microscopy, 15X magnification).}
    \item{\textbf{Movie S3:} \textbf{Non-flocking phase.} Time series of 1.5 µm-diameter beads propelled by an actin comet (phase contrast microscopy, 15X magnification, total duration: 5h).}
    \item{\textbf{Movie S4:} \textbf{Binary collision in the non-flocking phase}: 1.5 µm-diameter beads are propelled by an actin comet (phase contrast microscopy, 15X magnification).}
    \item{\textbf{Movie S5:} \textbf{Simulation of spontaneous symmetry breaking of a single bead.} In contrast to Movies S6 and S7, we let the bead naturally initiate motion. $\tilde{R} = 12.5$, $\tilde{k_1}=1.76$,$\tilde{k_2}=59.625$, and $\tilde{D_m}=1$, all other parameters are defined in Table \ref{tab:simulationparameters}}.    
    \item{\textbf{Movie S6:} \textbf{Simulation of single bead motility for varying radii} $\tilde{R}=\{10,15,20,25\}$, $\tilde{\Dm}=30$, and $\tilde{k}_2=3.3$, all other parameters are defined in Table \ref{tab:simulationparameters}. Columns from left to right are phase field $\phi$, actin monomer $\tilde{\cm}$, and actin polymer $\tilde{\cp}.$ An initial spot of polymer is applied to the right of the bead to initiate motion quickly, avoiding transients associated with spontaneous symmetry breaking.}
    \item{\textbf{Movie S7:} 
    \textbf{Simulation of single bead motility for varying reaction rate} $\tilde{k}_2=\{3,5,10,20\}$, $\tilde{\Dm}=1.0$, and $\tilde{R}=20$, all other parameters are defined in Table \ref{tab:simulationparameters}. Columns from left to right are phase field $\phi$, actin monomer $\tilde{\cm}$, and actin polymer $\tilde{\cp}.$ An initial spot of polymer is applied to the right of the bead to initiate motion quickly, avoiding transients associated with spontaneous symmetry breaking.}
    \item{\textbf{Movie S8:} \textbf{Simulation of multiple beads in the flocking state. }$\tilde{R}=20$, $\tilde{k_2}=10.0$, and $\tilde{\Dm}=10.0$, all other parameters are defined in Table \ref{tab:simulationparameters}}.
    \item{\textbf{Movie S9:} \textbf{Simulation of multiple beads in the non-flocking state.} $\tilde{R}=15$,$\tilde{k_2}=10.0$, and $\tilde{\Dm}=50.0$, all other parameters are defined in Table \ref{tab:simulationparameters}}.
\end{itemize}

\section{Materials and Methods}

\subsection{Protein Purification}
All recombinant proteins were expressed in BL21(DE3) cells from New England Biolabs. Final concentrations were measured using a linearized Bradford colorimetric assay with Bovine Serum Albumin (BSA) as a reference.

\subsubsection{Skeletal muscle actin from rabbit muscle acetone powder}
Rabbit skeletal muscle actin is purified following a protocol from Bruce Goode’s lab at Brandeis University based on Spudich and Watt 
\cite{Spudich1971_actin} with the following modifications. After centrifugation of the polymerized actin filaments, we resuspend the pellets in 30 mL of G-buffer (5 mM Tris HCl pH 8.0, 0.5 mM DTT, 0.2 mM ATP, 0.1 CaCl$_{2}$) supplemented with 10 mM dithiothreitol (DTT) and add them to a dounce homogenizer. The solution is dounced on ice until homogeneous. After incubating on ice for 1 hour, the solution is dounced once more before transferring into 10 kDa MWCO dialysis tubing. The sample is dialyzed for 2 days against G-buffer after which aggregates are removed by centrifugation at 265,000 x g for 30 minutes at 4\degree C. The final sample is then drop frozen, aliquoted, then stored at -80\degree C. When needed, an aliquot is removed from storage and allowed to thaw on ice overnight for use the following day.

\subsubsection{Capping protein}
\textit{Mus Musculus} CapZ is purified using a DNA plasmid supplied by Antoine J\'egou and Guillaume Romet-Lemonne (Addegene plasmid \# 89950)\cite{Shekhar2015_capzplasmid}. The original protocol \cite{Ulrichs2023_capz} utilized Ni-NTA beads which are tranferred to a disposable column, but here we use an \"{A}KTA FPLC. After, the cells are lysed in buffer (40 mM NaH$_{2}$PO$_{4}$ pH 7.4, 500 mM NaCl, 30 mM Imidazole, 1 mM DTT, and 0.1 mM EDTA) supplemented with protease inhibitor cocktail (PIC) tablet (ThermoFisher, \# A32955), DNase (New England Biolabs, M0303S), Phenylmethylsulfonyl Fluoride (PMSF, Millipore Sigma \# 52332-5GM), and lysozyme (US Biologicals \# L9200). Lysate is clarified and applied to a commercial 5 mL His-trap column (Cytiva) and eluted with lysis buffer with 500 mM imidazole. Protein is stored at -80\degree C in HEK buffer (20 mM HEPES pH 7.4, 1 mM EDTA, 50 mM KCl) with 10 mM DTT and 10\% glycerol. 

\subsubsection{Srv2}
His-tagged Srv2 is purified with nickel column affinity with the following modifications to the original protocol \cite{QuinteroMonzon2009_srv2}. After applying the cell lysate to the nickel column, 5 column volumes of wash buffer containing 50 mM NaH$_{2}$PO$_{4}$ pH 8.0, 300 mM NaCl, 140 mM Imidazole, and 1 mM DTT is flowed over the column. Srv2 is eluted using a linear gradient from 140 mM to 250 mM imidazole. Pooled fractions are dialyzed against 20 mM Tris HCl pH 8.0, 150 mM NaCl, and 2 mM DTT overnight then drop frozen and stored at -80\degree C. 

\subsubsection{Human Profilin-1}
Human profilin-1 is purified by affinity to poly-L-proline agarose \cite{Kaiser1989_pfn1}. The protocol followed is adapted from the original source above by the Mullins lab at the University of California San Francisco \cite{ucsfProfilinPurification}. Cell lysate is applied to a homemade poly-L-proline sepharose column (following an online protocol written by the previously mentioned Mullins lab which follows the above source) and eluted by denaturation by 6 M urea. Pooled fractions are step-dialyzed against decreasing concentrations of urea to renature the protein. The final sample is drop frozen and stored at -80\degree C.

\subsubsection{Human Cofilin-1}
Recombinant untagged human cofilin-1 is purified by closely following  \cite{Jansen2015_cfn1} via cation and anion exchange chromatography.

\subsubsection{GFP-pVCA}
The VCA domain of the Whiskott aldrich syndrome protein is expressed tagged with a GFP molecule in BL21 DE3 cells. The plasmid also includes a His-tag. Expressed cells are lysed in lysis buffer (50 mM NaH$_{2}$PO$_{4}$, 300 mM NaCl, 10 mM Imidazole, and 5\% glycerol) supplemented with 1 mM PMSF, $\frac{1}{4}$ PIC tab, 15 uL DNase, 5mg/ml lysozyme, 2 mM DTT, 0.5\% triton X-100, and 1 mM MgCl$_{2}$. After clarification by centrifugation, the sample is applied to a HisTrap HP (Cytiva), washed with lysis buffer, then eluted with a linear gradient of elution buffer (lysis buffer + 500 mM imidazole). Protein is pooled, dialyzed against HEK buffer supplemented with 10\% glycerol and 10 mM DTT. The sample is then aliquoted, flash frozen, and stored at -80\degree C.

\subsection{GFP-pVCA coated polystyrene microspheres}
Carboxylated polystyrene microspheres from PolySciences (6 $\mu m$ beads at 2.6\% solids cat\# 07312-5, 3 $\mu m$ beads at 2.7\% solids cat\# 17134-15, 1.5 $\mu m$ beads at 2.7\% solids cat\# 17133-15 ) are washed in HEK buffer 3 times (70 $\mu L$ bead stock + 30 $\mu L$ HEK) by centrifugation and resuspension. After, 20 uL of this mixture is suspended in 14 $\mu M$ GFP-pVCA, mixed well, and incubated on ice for 1 hour. The beads are pelleted using a small tabletop centrifuge and the supernatant is removed then replaced with HEK + 5 mg/ml BSA. This mixture is incubated for 15 minutes on ice. Then, in a similar fashion, the supernatant is replaced with HEK + 1 mg/ml BSA for storage at $4\degree \rm C$. Beads are used within 1 week of being made. 

\subsection{Flocking experiments}
To make an active sample of self-propelled beads, two separate tubes are prepared. All the stock and working concentrations are listed in \ref{tab:proteinconc}. The first tube contains G-buffer, G-actin, and exchange buffer (10x stock: 10 mM EGTA, 1 mM MgCl$_{2}$), which is mixed well before incubating at room temperature for at least 2 minutes. BSA and methylcellulose are added after incubation. A second tube is made with Arp2/3, profilin, cofilin, Srv2, capZ, and GFP-VCA coated beads. 

The two tubes are mixed together and 20x initiation mix (40 mM MgCl$_{2}$, 10 mM ATP, 1 M KCl) is added to initiate polymerization. After mixing the sample well, the desired volume is pipetted onto an acrylamide coated glass slide \cite{Sanchez2012_acryl} and gently covered with a cover slip. We controled the height of the sample chamber by varying the sample volume. The cmaber is sealed with either UV glue or 2 part epoxy. After curing, the sample is immediately taken for imaging. 


\begin{table}[h!]
    \centering
\begin{tabular}{||c c c||} 
 \hline
 Reagent & Stock Concentration ($\mu M$) & Working Concentration ($\mu M$) \\ [0.5ex] 
 \hline\hline
 Actin & 130 & 32 \\ 
 \hline
 Arp2/3 & 20 & 0.213 \\
 \hline
 Profilin & 177 & 12.5 \\
 \hline
 Cofilin & 163 & 22 \\
 \hline
 CapZ & 166 & 0.187 \\
 \hline
 GFP-pVCA & 33 & * \\
 \hline
 SrV-2 & 4.4 & 0.262 \\  
 \hline
 BSA & 100 mg/ml & 5 mg/ml \\ 
 \hline
 Methylcellulose & 2\% (w/w) & 0.03\% \\ [1ex]
 \hline
 
\end{tabular}
    \caption{\textbf{Protein stocks and working concentrations}. Working concentrations unless otherwise noted in figures and videos. *See above protocol for coating microspheres in GFP-pVCA.}
    \label{tab:proteinconc}
\end{table}

\subsection{Viscosity experiment}
Viscous media was prepared with PEO (molecular weight, 400,000 g/mol; Sigma-Aldrich) to the protein mix at 0.8$\%$ (w/v), which remain Newtonian (overlap concentration c*=3$\%$ (w/v)). PEO solutions were mixed for 2h at 40 r.p.m. (tube rotator; Fisherbrand). Viscosity measurements for the same polymer mixture were previously reported in \cite{stehnach2021viscophobic,ebagninin2009rheological}. 

\subsection{Bead-boundary experiment}
We designed a two-layered microfluidic device to study the interactions between motile beads and a either continuous or porous boundary.
Channels for the flow chamber are cut in a thin PDMS membrane (150-200 $\mu m$ thick). The reservoir channel was madedby casting PDM onto a 3D printed mold (Formlabs 3+ Resin Printer). A polyester membrane with a 0.8 um pore size (Steriltech Corporation, \#PET0847100) is placed over the PDMS piece such that the PDMS forms the bottom wall of the reservoir and the membrane is cut to size. For the non-porous boundary, the protein reservoir are replaced by solid block of PDMS. The device is sealed using UV glue (Norland Optical, part 8101). Once the device is assembled, two mixtures of proteins, one with pVCA-coated beads and one without, are prepared to be loaded into the device. The sample containing beads is manually loaded into the bottom sample chamber using a syringe and needle, while the sample without beads is manually loaded into the reservoir above using a syringe and Tygon Tubing (Cole-Parmer, ID 0.01 in, OD 0.03 in, \#06419-00) through the pre-punched holes on the top of the reservoir. The entire perimeter of the device including any inlets/outlets are then sealed using 2-part epoxy. A second experiment is run in parallel substituting the PDMS reservoir chip with the solid PDMS block. The filled devices are immediately taken for imaging. A  control experiment is performed simultaneously in quasi-2D flow chamber to confirm that the beads are motile and flocking in 2D.

\subsection{Microscopy}
Phase contrast and fluorescence images are taken with a Nikon TE2000 inverted microscope equipped with an Andor Zyla camera. Phase-contrast images were taken at 1 or 2 images per minute with either 10x or 20x objectives with a Ph1 phase ring. High-magnification fluorescence images of the actin were taken using a 100x oil immersion objective.
Data in the thick flow chambers (Fig 5) was collected using a laser scanning confocal microscope (TCSSP8, Leica Microsystems GmbH) where a 488 nm laser is used to excite GFP attached to the VCA on the surface of our polystyrene beads, a 20x air objective and 3 $\mu m$ z-step. Three positions with the porous membrane boundary plus 3 positions with a non-porous membrane boundary are scanned one after another with 10 minutes between scans.

\subsection{Data Analysis}
\subsubsection{Particle tracking}
The colloids were segmented and tracked with a subpixel accuracy ~\cite{Lu2007,Crocker1996,georgetownMatlabParticle}. We define the individual colloid velocities from their displacements over two subsequent frames with time interval $\delta t: v_i(t) = r_i(t + \delta t) - r_i(t)$, where $r_i(t)$ and $v_i(t)$ are respectively the position and velocity of a particle $i$ at time $t$.

\subsubsection{Criteria to detect flocks}
We define a flock as a group of at least two colloids that are close to each other and move in the same direction. In practice, we apply the following criteria to determine if two colloids $i$ and $j$ are flocking at time $t$: 
\begin{itemize}
    \item The distance between their center of mass is less than five times their radius $R: |r_i(t) - r_j(t)| < 5R$
    \item The difference between the orientation of their displacement is smaller than 45\degree: $| \theta_{j}(t)-\theta_{i}(t)| < \frac{\pi}{4}$, where $\theta_{i}(t)$ is the angle of the velocity $v_i(t)$. 
\end{itemize}

Two colloids verifying those criteria are called neighbors. We extract the list of neighbors for each colloid and reconstruct the flocks by applying the rule: two colloids sharing the same neighbors belong to the same flock. We then assign each colloid $i$ the size $n_{i}(t)$ of the flock it belongs to. $n_{i}(t)=1$ for single colloids.  

\subsubsection{Flocking parameter}
The instantaneous flocking parameter is defined as the ratio between the number of colloids belonging to a flock ($n_{i}(t) > 1$) and the total number of colloids in the field of view, at time $t$. The flocking parameter for a given experiment is the time average of the instantaneous flocking parameter. \\

\subsubsection{Analysis of confocal data}
pVCA-covered Beads in confocal image stacks are segmented using MATLAB. A threshold is applied to each confocal image, and detected objects are filtered by size such that small objects of only a few pixels are removed. Then, the circularity of each object is computed to remove irregular objects and select only the fluorescent beads in the field of view. The bead density is measured by integrating the number of beads per z-slice.

\subsubsection{Measurement of exponential decay length $\xi$ from single-bead simulations}
Simulations of single motile beads reveals the spatial distribution of actin monomers around a motile bead. Such profiles are shown in Fig. 2F,G of the main text. We measured the decay length of the actin monomer concentration along the direction perpendicular to the beads' motion. The actin monomer profiled is well described by a double exponential. The decay length, $\xi$, corresponds to the shortest of the two exponential length. The concentration profile is fitted using a non-linear squares method.

\subsubsection{Statistical analysis and number of independent replicates}
When appropriate, mean and standard deviations of the experimental data are shown as a disk and an errorbar (Insert of Fig. 2B and 2C). In Fig. 4H,I,J, each dot represents the time average flocking parameter for an independent experiment.

\section{Model}

We model the dynamics of polymerization-driven beads through coupled PDEs describing reaction-diffusion processes, active hydrodynamics, and phase fields for capturing bead dynamics.  In this section, we first derive the model and then non-dimensionalize it using energy, time, and length scales intrinsic to the equations.


\subsection{Free energy potentials for phase field and reactive species}
A Cahn-Hilliard free energy with minima at $\phi_{\pm}=\{0,1\}$ is used to define particle bodies

\begin{equation}
f_{\phi_i}=\kappa \frac{1}{2}\phi_i^2 \left(1-\phi\right)^2 + \frac{1}{2}\gamma \nabla\phi_i\cdot\nabla\phi_i+
\kappa_{\phi\phi}\sum_{j=1,i\neq j}^N\frac{1}{2}\phi_i^2 \phi_j^2
+\lambda A_0 \frac{1}{2}\left(1-\frac{1}{A_0}
    \int\phi_i^2 dx
    \right)^2
,
\end{equation}
where the index indicates the $i^{\text{th}}$ particle. The first two terms are the standard Cahn-Hilliard terms responsible for setting the equilibrium dense and dilute values of the field $\phi$ and the interface width. The third term describes steric repulsion between beads, and the final term is a Lagrange multiplier that maintains uniform bead size by penalizing deviations from  $A_0$ For sufficiently large $\lambda$. To approximate the impermeability of the bead surface, we include free energy terms that promote cross-diffusion between the chemical species and the phase field variables
\begin{equation}
    f_{c}=\kappa_c\frac{1}{2}\left(\cm^2+\cp^2\right)\left(1+\sum_{i=1}^N\phi_i^2\right).
    \label{eq:phi_dim} 
\end{equation}
We evolve the phase field using non-conserved gradient descent dynamics, known in the literature as model A,
\begin{align}
        \frac{\partial \phi_i}{\partial t}+\bu\cdot\nabla \phi_i &= -\frac{1}{\Gamma} \frac{\delta F_{\phi}}{\delta \phi_i},        
\end{align}
where $F_\phi = \int  f_{\phi} dx$ and
\begin{equation}
    \frac{\delta F_{\phi}}{\delta\phi_i} = \kappa\left(2\phi_i^3-\phi_i\right) + \gamma \nabla^2 \phi_i + \kappa_{\phi\phi} \phi_i \sum_{j=1,i\neq j}^N
     \left[
    \phi_j^2 + \lambda \phi_i \left(1-\frac{1}{A_0}\int{\phi_i^2 dx} \right)
    \right].
\end{equation}

\subsection{Reaction-diffusion dynamics of monomer and polymer fields}
We model the polymerization reaction $R$ as a two-step process consisting of nucleation $\propto k_1$ and growth $\propto k_2$


\begin{equation}
    R_{\text{on}}\left(\cm,\cp\right)=\left(k_1 + k_2 \frac{\cp^2} {\kd^2 + \cp^2 } \right) \cm.
\end{equation}
We model the nonlinear growth using a Hill function. This form has been used in prior works to capture cooperative growth in polymerizing actin networks and helps stabilize the simulations by preventing run-away growth.

The density of monomers and polymers ($\cm$ and $\cp$) follow standard diffusive dynamics, with diffusion constants $\Dm$ and $\Dp$ respectively. They are also advected by the flow $\bu$, and react through the reaction term described above, modulated by the presence of interfaces ($|\nabla\phi|^2$). Polymers also decay back to monomers at a rate $\koff$. These dynamics are contained in the following equations





\begin{align}
    \frac{\partial \cm}{\partial t}+\bu\cdot\nabla \cm &= \Dm\nabla\cdot\nabla\frac{\delta F_c}{\delta \cm} - |\nabla\phi|^2 l^2_{\gamma}R_{\text{on}}+\koff \cp,
\\
    \frac{\partial \cp}{\partial t}+\bu\cdot\nabla \cp &= \Dp\nabla\cdot\nabla\frac{\delta F_c}{\delta \cp} + |\nabla\phi|^2 l^2_{\gamma} R_{\text{on}}-\koff \cp,
\end{align}
and
\begin{equation}
    \frac{\delta F_c}{\delta c_{\text{m,p}}} = 
    \kappac c_{\text{m,p}}
    \left(
    1 + \sum_{i=1}^N \phi_i^2
    \right).
\end{equation}


\subsection{Hydrodynamics of bead transport}

Particle motion is driven by forces generated at the surface through polymerization.  In the Stoke's limit with substrate friction, momentum balance requires
\begin{equation}
    \nabla\cdot\sigma + f_a -\xi u = 0,
\end{equation}
Where $\sigma = \sigma_{a} + \sigma_c + \sigma_u $. The capillary body force arising from the phase field interface curvature is given by
\begin{equation}
    \sigma_c = -\gamma\sum_{i=1}^N\nabla\phi_i\otimes\nabla\phi_i,
\end{equation}
where we've omitted terms that would only contribute to the static pressure and not materially impact the dynamics. Equivalently, $\nabla\cdot\sigma_c=\sum_{i=1}^N \phi_i\nabla\mu_i$.

The active stress from actin polymerization follows the same form as the passive surface tension and is given by
\begin{equation}
    \sigma_a =-\alpha l_{\gamma}^2 R_{\text{on}}\left(c_m,c_p\right) \sum_{i=1}^N\nabla\phi_i\otimes\nabla\phi_i,
\end{equation}
where $R_{\text{on}}$ is the total polymerization rate from before and $\alpha$ is the scale with the appropriate units to convert this rate into a stress. We include the factor $l_{\gamma}^2$ to account for the fact that the interface width impacts the total active stress.

For $\alpha > 0$, this stress is contractile along the interface. Thus, this term can be viewed as a form of active surface tension in the bead's actin cortex whose gradients will lead to material flows.  Alternatively, in this coarse-grained view, the active stress can be viewed as extensile with an axis defined by $\hat{e}_n=\nabla\phi/|\nabla\phi|$. In other words, a contractile cortex or an extensile tail are equally valid ways of interpreting this active stress. In either case, the surface normal sets the anisotropy axis of the stress. We note that similar representations of active stresses arising from surface-bound polymerization have been used in other phase field models of motility. For convenience,  all of the terms together give
\begin{equation}    \sum_{i=1}^N\nabla\cdot\left[-\left(\alpha l_{\gamma}^2\Ron\left(\cm,\cp\right)+\gamma\right)\nabla\phi_i\otimes\nabla\phi_i\right] +\nabla^2\bu -\xi \bu -\nabla P= 0.
\end{equation}

\begin{table}[h!]
    \centering
    \begin{tabular}{|c|c|c|}
        \hline
         parameter description & symbol & value  \\
         \hline
         dimensionless diffusivity of actin monomers $\cm$ & $\tilde{\Dm}$ & 1-100  \\
         dimensionless diffusivity of polymerized actin $\cp$ & $\tilde{\Dp}$ & 0.1  \\
         dimensionless nucleation rate  & $\tilde{k}_1$ & 0.1-2  \\
         dimensionless growth rate  & $\tilde{k}_2$ & 1-20  \\
         dimensionless disassociation constant for Hill equation & $\tilde{\kd}$ & 0.3  \\
        dimensionless depolymerization rate & $\tilde{\koff}$ & 0.006 \\
         dimensionless polymerization stress factor & $\tilde{\alpha}$ & 5-30 \\
         ratio of phase field interface thickness to hydrodynamic screening length & $\epsilon$ & $0.2$ \\
         dimensionless bead radius & $\tilde{R}$ & 10-40 \\
         \hline
        spatial domain size & $L_{\text{X}} \times L_{\text{Y}}$ & $256^2$ - $1024^2$  \\
        number of spatial modes & $N_{\text{X}} \times N_{\text{Y}}$ & $256^2$ - $1024^2$ \\
        typical time step size & $\Delta t$ & 0.05 \\
        typical simulation duration & $T$ & $10^6 \Delta t$ \\ 
         \hline
    \end{tabular}
    \caption{Dimensionless parameters for numerical simulations.}
    \label{tab:simulationparameters}
\end{table}

\subsection{Dimensionless Equations}
In order to develop a reduced set of control parameters, we non-dimensionalize our system of equations. The coupled reaction-diffusion-convection equations possess numerous time and length scales. We choose the length scale $l\sim\sqrt{\gamma/\kappa}$, which is approximately the interface width, and for the time scale, the dissipation rate from the phase field dynamics $t\sim\Gamma/\kappa$. For both $\cm$ and $\cp$, we use the initial concentration $\cmIC$ as the concentration scale. Applying these scales to \eqn\ref{eq:phi_dim} yields the dimensionless phase field dynamics:
\begin{equation}
        \frac{\partial \phi_i}{\partial \tilde{t}}+\tilde{\bu}\cdot\tilde{\nabla} \phi_i =
        -\phi_i\left(1-3\phi_i + 2\phi_i^2\right) +
        \tilde{\nabla}^2\phi_i +
       \tilde{\kappa}_{\phi\phi}\sum_{j=1,i\neq j}^N \phi_i \phi_j^2+
        \lambda \phi_i \left(1-\frac{1}{\tilde{A}_0}\int{\phi_i^2 d\tilde{x}} \right),
        \label{eq:phi_nondim}
\end{equation}
where $\tilde{A}_0=\pi \frac{R^2}{l_{\gamma}^2}$ is the dimensionless area of a single particle and $ \tilde{\kappa}_{\phi\phi}=\frac{\kappa_{\phi\phi}}{\kappa}$ is just a ratio of energy constants. The dimensionless dynamics for monomer and polymer fields are then
\begin{align}
    \frac{\partial \tilde{\cm}}{\partial \tilde{t}}+\tilde{\bu}\cdot\tilde{\nabla} \cm &= \tilde{\Dm}\tilde{\nabla}^2 \left(\cm\left(1+\sum_{i=1}^N\phi_i^2\right)\right) - |\tilde{\nabla}\phi|^2 \tilde{R}_{\text{on}}+\tilde{\koff} \tilde{\cp},
    \label{eq:cm_nondim}
    \\
    \frac{\partial \tilde{\cp}}{\partial \tilde{t}}+\tilde{\bu}\cdot\tilde{\nabla} \cp &= \tilde{\Dp}\tilde{\nabla}^2 \left(\cp\left(1+\sum_{i=1}^N\phi_i^2\right)\right)+|\tilde{\nabla}\phi|^2 \tilde{R}_{\text{on}}-\tilde{\koff} \tilde{\cp},
    \label{eq:cp_nondim}
    \\
\end{align}
where $\tilde{D}=\frac{D \Gamma \kappac}{\gamma}$, $\tilde{\koff}=\frac{\koff \Gamma}{\kappa}$. In experiments, the Peclet number for actin monomers $\text{Pe}_{\text{m}}\ll1$. For numerical convenience, we strictly enforce this limit by turning off convection in \eqn \ref{eq:cm_nondim} and \eqn\ref{eq:cp_nondim} (Note that convection of the bead remains on). The dimensionless reaction rate is

\begin{equation}
    \tilde{R}_{\text{on}} = \left(
    \tilde{k}_1+\tilde{k}_2\frac{\tilde{\cp}^2}{\tilde{\kd}^2+\tilde{\cp}^2}\right) \tilde{\cm},
    \label{eq:stokes_nondim}
\end{equation}
with $\tilde{k}_1=k_1 \frac{\Gamma}{\kappa}$, $\tilde{k}_2=k_2 \frac{\Gamma}{\kappa}$, and $\tilde{\kd}=\kd /\cmIC$. Finally, the dimensionless Stokes momentum equation is given by
\begin{equation}    \sum_{i=1}^N\tilde{\nabla}\cdot\left[-\left(
\tilde{\alpha}\tilde{\Ron}+\tilde{\Gamma}\right)\tilde{\nabla}\phi_i\otimes\tilde{\nabla}\phi_i\right] +\tilde{\nabla}^2\tilde{\bu} -\epsilon^2 \tilde{\bu} -\tilde{\nabla} \tilde{P}= 0,
\end{equation}
where $\tilde{\alpha}=\frac{\alpha\cmIC}{\eta}$ is the dimensionless polymerization stress, $\tilde{\Gamma}=\frac{\Gamma}{\eta}$ a ratio of dissipation rates, and $\epsilon=\sqrt{\frac{\gamma}{\kappa}}\sqrt{\frac{\xi}{\eta}} $ is a dimensionless ratio between the interface thickness and the hydrodynamic screening length $\sim\sqrt{\eta/\xi}$. 

\subsection{Numerical details}

We have integrated \eqns\ref{eq:phi_nondim}-\ref{eq:stokes_nondim}  numerically using a pseudo-spectral method, with standard Euler-Maruyama time stepping. We have used a publicly available software package to do so, cuPSS \cite{caballero2024} and the code is available on github. The parameters used in the simulation are shown in Table \ref{tab:simulationparameters}.

\section{Testing other mechanisms that could lead to flocking}

We considered two other possible mechanisms for flocking: hydrodynamic interactions \cite{TONER2005170} and motility-induced phase separation \cite{Cates2015}, but quickly ruled them out. We eliminated hydrodynamic interactions as the dominant interaction mode by examining the motion of small passive tracer particles in the vicinity of a motile active bead.  We noted that their motion was dominated by diffusion and did not measure any hydrodynamic flows around motile active beads. We further ruled out motility-induced phase separation as long-lived flocks with only two beads were routinely observed, even in the dilute regime.

\section{Supplementary figures}

\begin{figure}[h!]
    \centering
    \includegraphics[width=0.75\linewidth]{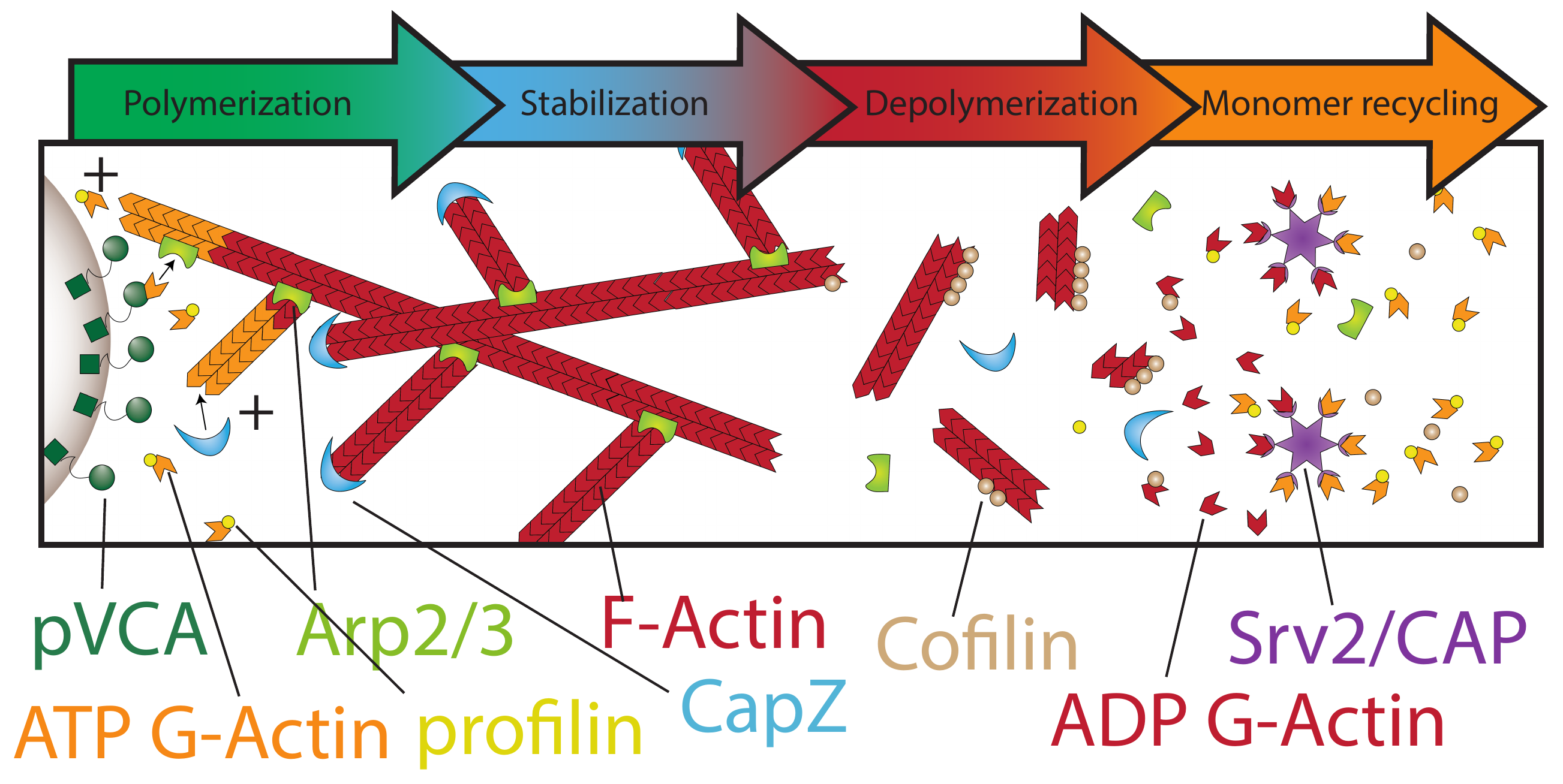}
    \caption{\textbf{Schematic of the biochemical reactions that drive the motion of the bead.} The bead surface, covered with pVCA, is on the left of the schematic.}
    \label{figS1}
\end{figure}

\begin{figure}[h!]
    \centering
    \includegraphics[width=0.75\linewidth]{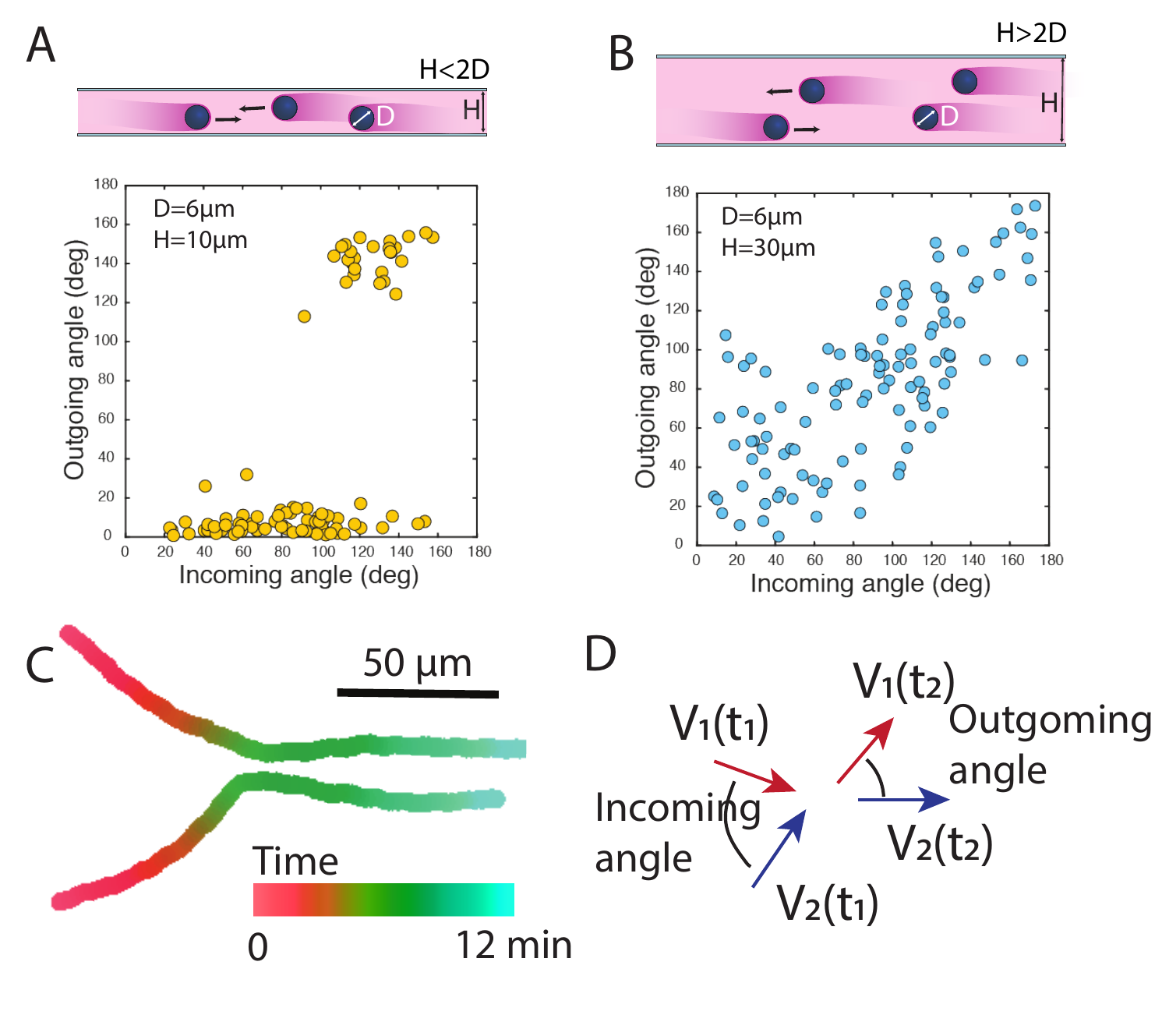}
    \caption{\textbf{Quasi-2D confinement is necessary for flocking.} A) Plot of the velocity angles before and after collision for beads in quasi-2D chambers. Motile beads confined in quasi-2D chambers interact sterically and reorient after colliding. B) Plot of the velocity angles before and after collision for beads in 3D chambers. Motile beads in thicker chambers are able to move on top of each other. As a result, they do not reorient after colliding. C) Typical trajectory associated with panel A (beads in quasi-2D confinement, Movie S2). D) Definition of the incoming and outgoing angles.}
    \label{fig:enter-label}
\end{figure}

\begin{figure}[h!]
    \centering
    \includegraphics[width=0.5\linewidth]{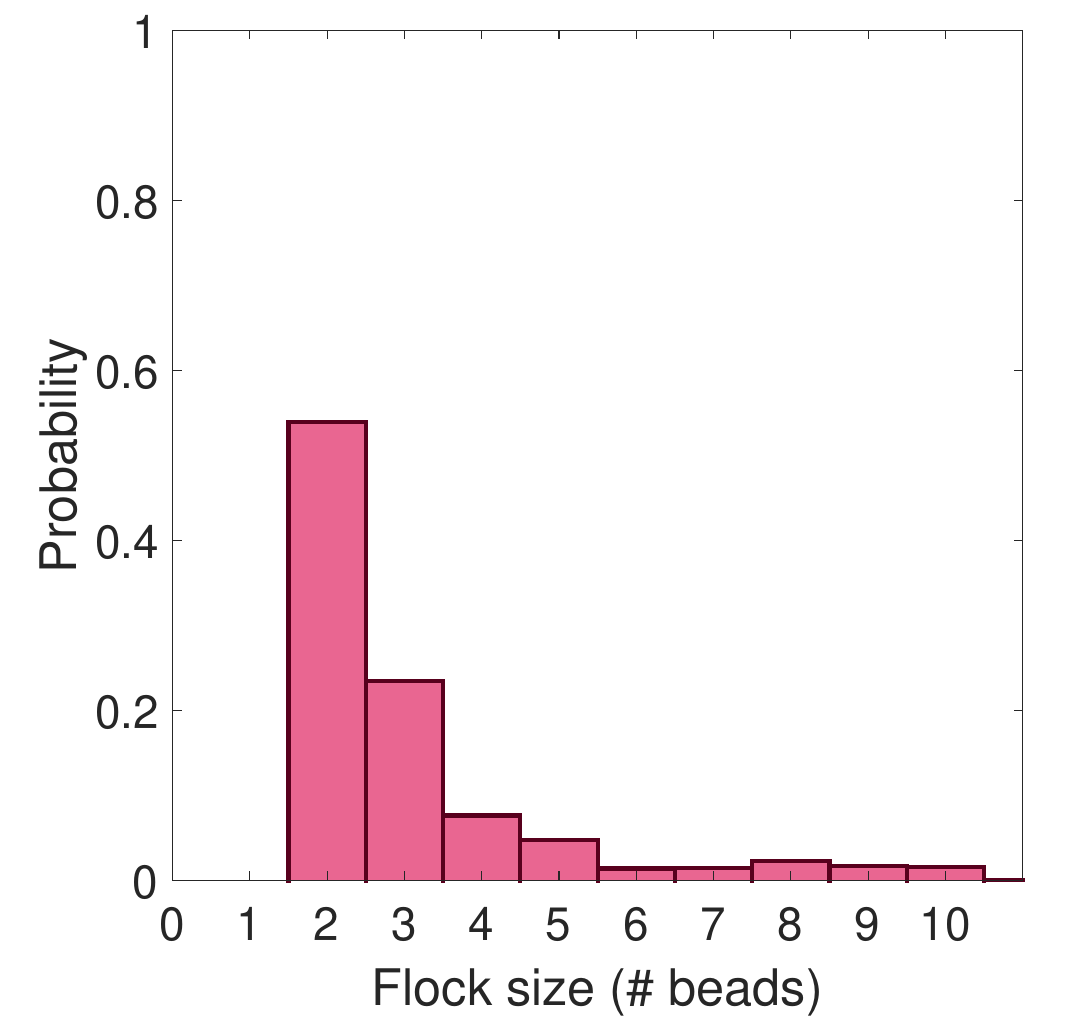}
    \caption{\textbf{Probability distribution of the size of the flocks.} Total number of flocks analyzed ($n\ge 2$): 16890 (over 15 independent experiments, with 1000 images per experiment, beads $6\, \rm \mu m$, starting $125\, \rm min$ after beginning of the movies to ensure steady state regime).}
    \label{fig:SI_PDF_sizeFlocks}
\end{figure}

\begin{figure}[h!]
    \centering
    \includegraphics[width=1\linewidth]{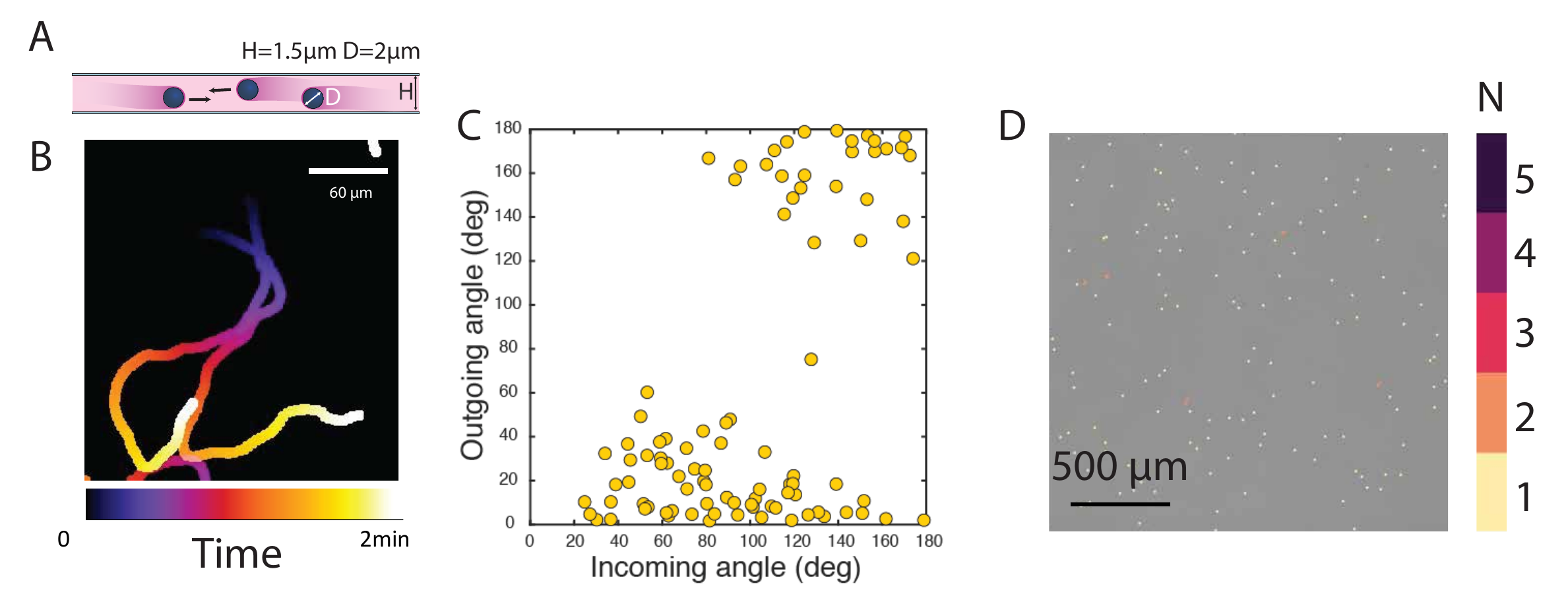}
    \caption{\textbf{Small beads in quasi-2D chambers interact sterically but do not flock.} A) schematic of a side view of the experiment. B) time-colored trajectories of two colliding beads (Movie S4) C) Plot of the velocities angle just before and just after collision for N=100 binary collisions. D) Snapshot of an experiment where beads are color coded according to the size of the flock they belong to. Most of the beads move individually. }
    \label{fig:SI_PDF_sizeFlocks}
\end{figure}

\newpage
\begin{figure}[h!]
    \centering
    \includegraphics[width=1\linewidth]{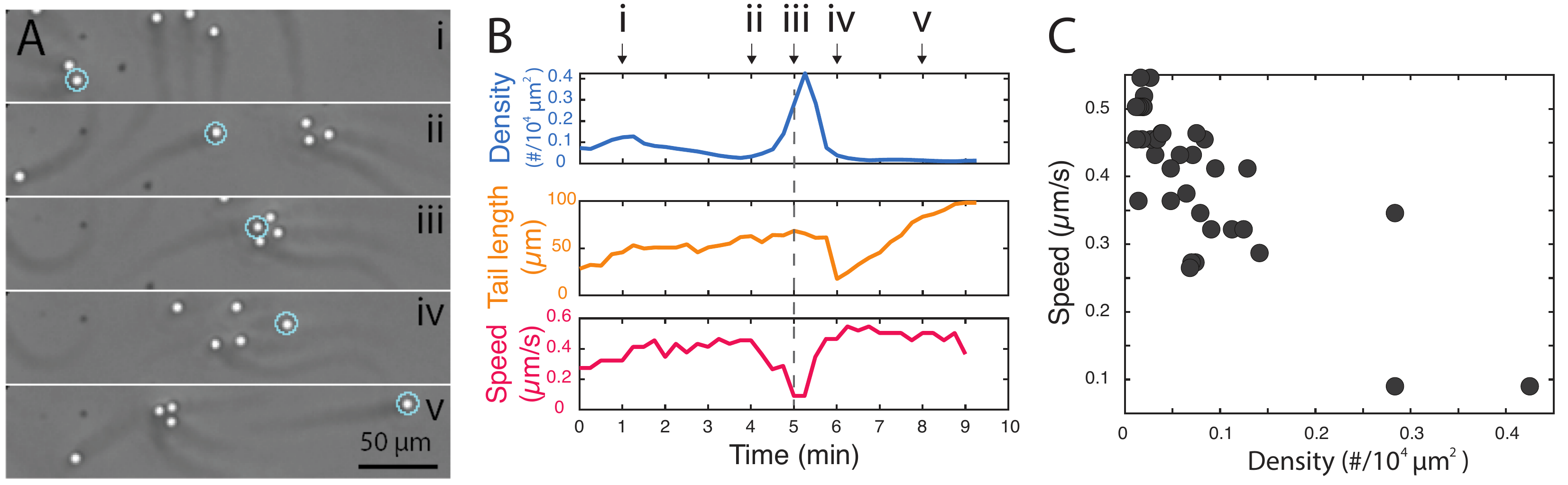}
    \caption{\textbf{Bead speed, tail length, and tail density are impacted by local bead density.} A) Time series of a bead, marked by cyan circle, moving through a 3-beads flock. B) Time evolution of the local bead density, tail length and bead speed. C) Plot of bead speed as a function of local bead density. Beads slow down as the density increases.}
    \label{fig:enter-label}
\end{figure}

\begin{figure}[h!]
    \centering
    \includegraphics[width=0.75\linewidth]{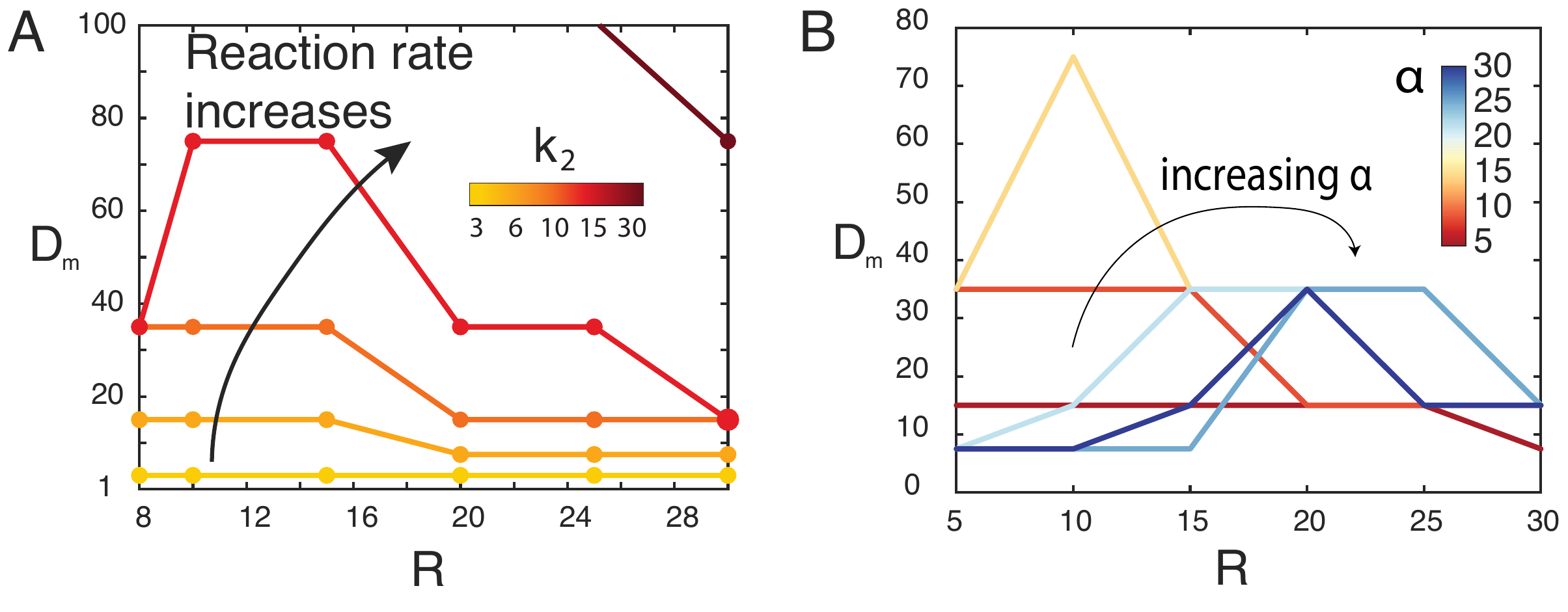}
    \caption{\textbf{Impact of the reaction rate $k_2$ and the activity $\alpha$ on the flocking transition.} A-B) Phase boundary in the $(D_m,R)$ phase space where A) $k_2$ increases and B) $\alpha$ increases. $D_m$ is the diffusivity of actin monomers, R is the beads radius. The phase boundary is defined by the midpoint between a flocking and non-flocking parameter set for a given range of radii values. }
    \label{figS7}
\end{figure}

\begin{figure}[h!]
    \centering
    \includegraphics[width=0.75\linewidth]{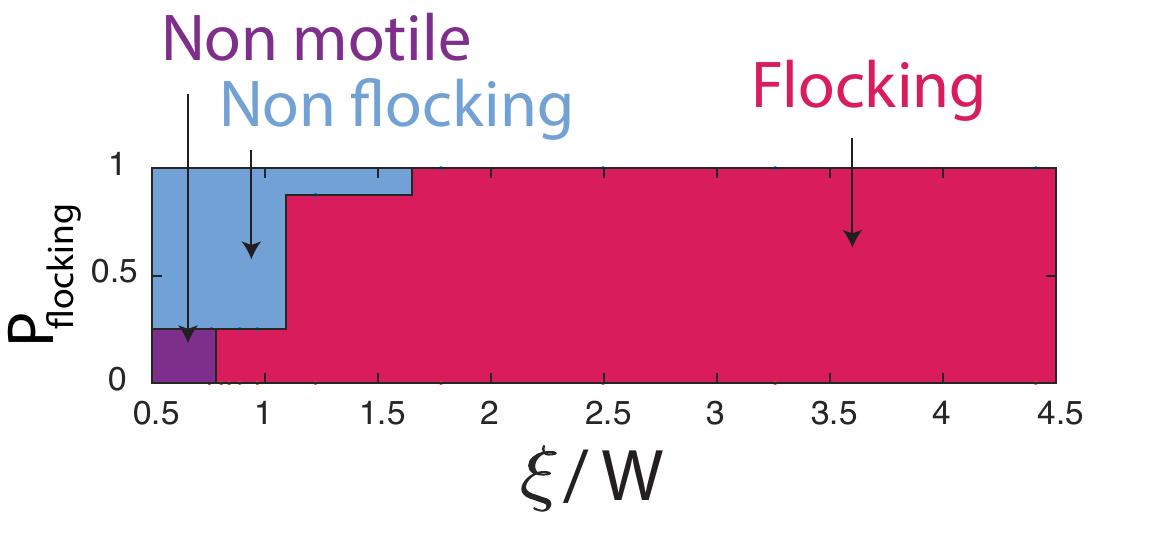}
    \caption{\textbf{Phase behaviors as a function of the decay length of the actin monomer gradient $\xi$ and W, the thickness of the phase field boundary.} We measured $\xi$ from simulations of single beads. The emergent dynamics are determined by inspecting multi-bead simulations with the same parameters.}
    \label{figS7}
\end{figure}

\begin{figure}[h!]
    \centering
    \includegraphics[width=0.75\linewidth]{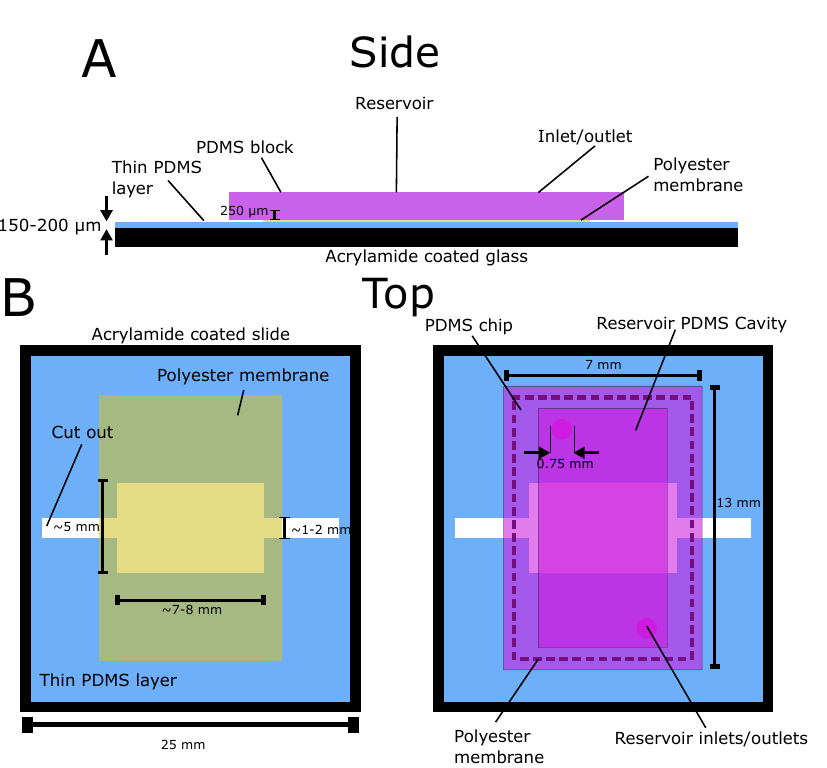}
    \caption{\textbf{Schematic of the two-layer microfluidic device for boundary experiments.} Schematic not to scale. A) Side view of the device. The sample chamber is in the bottom blue layer while the protein reservoir is in the magenta layer. The two layers are separated by a porous membrane. Dotted lines represent the inlet and outlets in the top layer. B-Left) Top view of the sample chamber (bottom layer). A channel is cut out of a thin PDMS layer on top of a glass slide. Just above this cut out is the polyester membrane which forms the ceiling of the chamber below. The two smaller channels serve as the inlet/outlet pair to fill this chamber. B-Right) Top view of the protein reservoir (top chamber). A PDMS block with a rectangular inset sits atop the membrane such that the membrane makes up the sixth wall of the chamber. Two holes are punched through the top of the block which are the inlet/outlet pair to fill the reservoir.}
    \label{fig:enter-label}
\end{figure}

\clearpage

\bibliographystyle{apsrev4-1}
\bibliography{main}

\end{document}